\documentclass{JHEP}

\usepackage{graphicx}
\usepackage{epsfig}
\usepackage{amsfonts}
\usepackage{amssymb}
\usepackage{amsmath}

\newcommand{\cgg}[2]{{#2}}

\def\be{\begin{equation}}
\def\ee{\end{equation}}

\newcommand{\bea}{\begin{eqnarray}}
\newcommand{\eea}{\end{eqnarray}}
\newcommand{\bean}{\begin{eqnarray*}}
\newcommand{\eean}{\end{eqnarray*}}
\newcommand{\nn}{\nonumber \\}

\def\W #1{\widetilde{#1}}

\def\Tr{\mathop{\rm Tr}}

\def\Sl{\sum\limits}
\def\Label#1{\label{#1}%
  \smash{\hbox to0pt{\raise1ex\hbox{\tiny[#1]}\hss}}}

\newcommand{\ctobedelete}[1]{}
\allowdisplaybreaks
\title{Amplitude Relations in Non-linear Sigma Model}

\author{Gang Chen${}^{a}$, Yi-Jian Du${}^{b,c}$\footnote{Corresponding author} \\
$^a$\small Department of Physics, Nanjing University 22 Hankou Road, Nanjing 210093, China\\
$^b$\small Department of Physics and Center for Field Theory and Particle Physics, Fudan University, \\
~220 Handan Road, Shanghai 200433, P.R China\\
$^c$\small Department of Physics and Astronomy, University of Utah, Salt Lake City, UT 84112, USA\\
~~~~~~~\\
\email{gang.chern@gmail.com; yjdu@fudan.edu.cn\hskip0.5cm}}

\date{\today}
\abstract{In this paper, we investigate tree-level scattering amplitude relations in $U(N)$ non-linear sigma model.
We use Cayley parametrization. As was shown in the recent works \cite{Kampf:2012fn,Kampf:2013vha}, both on-shell amplitudes and off-shell currents with odd points have to vanish under Cayley parametrization. We prove the off-shell $U(1)$ identity and fundamental BCJ relation for even-point currents. By taking the on-shell limits of the off-shell relations, we show that the color-ordered tree amplitudes with even points satisfy $U(1)$-decoupling identity and fundamental BCJ relation, which have the same formations within Yang-Mills theory. We further state that all the on-shell general KK, BCJ relations as well as the minimal-basis expansion are also satisfied by color-ordered tree amplitudes. As a consequence of the relations among color-ordered amplitudes, the total $2m$-point tree amplitudes satisfy DDM form of color decomposition as well as KLT relation.
}


\begin{document}

\section{Introduction}

One of the most significant progresses of scattering amplitudes in recent years is the discovery of new amplitude relations.
The new relation (BCJ relation) was firstly proposed in Yang-Mills theory by Bern, Carrasco and Johansson\cite{Bern:2008qj}. Using BCJ relation in addition with
KK relation which was earlier suggested by Kleiss and Kuijf \cite{Kleiss:1988ne}, one can simplify the calculations on color-ordered amplitudes at tree level. In particular, these relations provide a reduction of the basis of $n$-point tree-level amplitudes to a minimal basis of $(n-3)!$ independent ones\cite{Kleiss:1988ne}. Tree-level amplitude relations in Yang-Mills theory have been studied in both string theory and field theory. In string theory, both KK and BCJ relations can be considered as so-called monodromy relations \cite{BjerrumBohr:2009rd,Stieberger:2009hq}. In field theory, KK relation was firstly proved via new color decompositions \cite{DelDuca:1999rs}, while, both KK and BCJ relations have been proved by BCFW recursion \cite{Britto:2004ap, Britto:2005fq}(the proof of KK relation and fundamental BCJ relation can be found in \cite{Feng:2010my}\footnote{Other approaches to fundamental BCJ relation can be found in \cite{Tye:2010kg}, \cite{Cachazo:2012uq}.}, the proof of general BCJ relation was given in \cite{Chen:2011jxa}). The minimal-basis expansion has been proved \cite{Chen:2011jxa} via so-called general BCJ relation.

KK and BCJ relations in Yang-Mills theory can be regarded as results of color-kinematic duality \cite{Bern:2008qj}. In \cite{Bern:2008qj}, it was pointed that one could express the amplitudes by Feynman-like diagrams with only cubic vertices and establish a duality between color factors and kinematic factors. Once the color factors satisfy some algebraic property (antisymmetry and Jacobi identity), so do the corresponding kinematic factors. In fact, KK relation among color-ordered amplitudes can be considered as a result of antisymmetry of kinematic factors, while, BCJ relation is a result of Jacobi identity. The kinematic factors in Yang-Mills theory can be constructed from pure spinor string theory \cite{Mafra:2011kj}. They can also be constructed by area-preserving diffeomorphism algebra \cite{Monteiro:2011pc,BjerrumBohr:2012mg} or a more general diffeomorphism algebra\cite{Fu:2012uy}. A further understanding of the kinematic algebra is the construction of color-dual decomposition and trace-like objects \cite{Bern:2011ia,Du:2013sha,Fu:2013qna}.

It is interesting that KK and BCJ relations can be found not only in pure Yang-Mills theory but also in other theories. For example, relations for amplitudes with gauge field coupled with matter was investigated in \cite{Sondergaard:2009za}. In $\mathcal{N}=4$ super Yang-Mills theory, the super-amplitudes are also proven to satisfy KK and BCJ relations \cite{Jia:2010nz}. In \cite{Du:2011js}, the KK and BCJ relations was proven to hold for color-scalar amplitudes. Though these amplitude relations are found in different theories, they have similar forms with the relations in Yang-Mills theory. This is because the color-kinematic duality implies that different theories with color factors satisfying the same algebraic properties should have the similar form of amplitude relations. When the algebraic properties are changed, the amplitude relations should also be changed. This can be further supported by the amplitude relations in three dimensional supper symmetric theory with 3-algebra \cite{Bargheer:2012gv}. In this case, the algebraic properties of color factors are changed to the properties of 3-algebra, the form of amplitude relations are also changed to agree with the algebraic structure.

Beyond the  fundamental field theory, there are lots of interesting low energy effective theories which are also widely used in the phenomenology of low energy physics. One of them is the well-known $SU(N)$ non-linear sigma model. This theory describes the low energy dynamics of the Goldstone Bosons under the chiral symmetry breaking $SU(N)_L\times SU(N)_R\rightarrow SU(N)$.  In this paper,  we focus on the relations of tree-level amplitudes in $U(N)$ non-linear sigma model. For on-shell amplitudes, the result can apply to the $SU(N)$ model directly. In recent works \cite{Kampf:2012fn,Kampf:2013vha}, $U(1)$-decoupling identity was discussed via the decoupling of $U(1)$ field from interaction, and color-order reversed relation was also pointed in \cite{Kampf:2013vha}. These results encourage us to study the full amplitude relations in non-linear sigma model systematically.
We expect that there should be KK and BCJ relations, which share the same forms with the relations in Yang-Mills theory.
This is because the color factors\footnote{Although, in non-linear sigma model, one may use \emph{flavor} factor instead of \emph{color} factor, as was done in \cite{Kampf:2013vha} for physical reason, we will use color through this paper for convenience. We hope this will not make any confusion.} in these two cases satisfy the same algebraic properties. However, the kinematic factors which share the same algebraic properties  cannot easy to construct because of the infinity of the number of vertices in non-linear sigma model. The general amplitude relations are also not obvious along the decoupling argument in \cite{Kampf:2012fn,Kampf:2013vha}. In fact, the arguments on $U(1)$-decoupling identity in \cite{Kampf:2012fn,Kampf:2013vha} are valid for only on-shell amplitudes. When we consider the even-point off-shell currents constructed by Feynman rules, the $U(1)$-field under Cayley parametrization \cite{Kampf:2012fn,Kampf:2013vha} do not decouple from interaction. This is quite different from in case of Yang-Mills theory where both on-shell amplitudes and off-shell currents satisfy KK relation (the KK relation in off-shell case in Yang-Mills theory was proven in the appendix of \cite{Fu:2012uy}). Furthermore, the highly nontrivial relations-BCJ relations seem hard to obtained from this argument.
One may hope to prove the relations by using the nontrivial extension of BCFW recursion in non-linear sigma model \cite{Kampf:2012fn,Kampf:2013vha} and follow the similar proof within Yang-Mills case \cite{Feng:2010my,Chen:2011jxa}, but it will be not easy to use the Even(odd)-shift form of the BCFW recursion \cite{Kampf:2012fn,Kampf:2013vha} to prove even if the simple case-$U(1)$ decoupling identity.

In this work, we will use Berends-Giele recursion\footnote{Berends-Giele recursion was firstly given in Yang-Mills theory in \cite{Berends:1987me}. The Berends-Giele recursion in non-linear sigma model was proposed in the recent work \cite{Kampf:2012fn,Kampf:2013vha}.} under Cayley parametrization to study the relations. Since the odd-point amplitudes vanish \cite{Kampf:2012fn,Kampf:2013vha}, we only need to study the relations for even-point amplitudes. We conjecture and prove $U(1)$ identity\footnote{In off-shell case, we use '$U(1)$ identity' instead of '$U(1)$-decoupling identity' because in the off-shell case, the $U(1)$ gauge field in general cannot decouple. Only in the on-shell case, the $U(1)$ gauge field decouples.} and fundamental BCJ relation for even-point off-shell currents. We will find that, the left hand side of the the $U(1)$ identity and fundamental BCJ relation must equal to terms proportional to $(p_1^2)^0$, where $p_1$ is the momentum of the off-shell leg. When we turn our attention to on-shell amplitudes, we should multiply the current by
 $p_1^2$ and take the on-shell limit $p_1^2\rightarrow 0$. Then we get the $U(1)$-decoupling identity and fundamental BCJ relations for on-shell amplitudes. We will leave the proof of general off-shell relations in future work.

Though it will be hard to derive off-shell general BCJ relation from either Berends-Giele recursion or BCFW recursion, \cite{Ma:2011um} provides another method to prove the general KK and BCJ relations. It was pointed out that all the on-shell general KK and general BCJ relations can be generated by the fundamental BCJ relation as well as cyclic symmetry. In non-linear sigma model, at on-shell case, both fundamental BCJ relation and cyclic symmetry are satisfied, thus we also have general KK and general BCJ relations. Since the general KK and BCJ relations are satisfied, consequent results such as minimal-basis expansion, Del Duca-Dixon-Maltoni(DDM) color decomposition \cite{DelDuca:1999rs} and the $(2n-2)!$-formula \cite{BjerrumBohr:2010ta} of   Kawai-Lewellen-Tye(KLT) relation \cite{Kawai:1985xq} for $2n$-point amplitudes can be derived.

The structure of this paper is following. In section 2, we provide a short review of Feynman rules and Berends-Giele recursion in non-linear sigma model. In section 3, we will prove the off-shell $U(1)$ identity. We first give some examples then the general proof. In section 4, we will prove the off-shell BCJ relation. We also give examples before general proof. After taking the on-shell limits of the off-shell KK and BCJ relations, we can obtain the $U(1)$-decoupling identity and fundamental BCJ relation for on-shell amplitudes immediately. In section 5, we use the conclusions of the work \cite{Ma:2011um} to state that all the on-shell general KK and BCJ relations can be generated by the on-shell fundamental BCJ relation as well as cyclic symmetry. Thus the on-shell general KK and general BCJ relations are naturally satisfied. We also point out that the minimal-basis expansion of color-ordered amplitudes, DDM color decomposition and the $(2m-2)!$ formula of KLT relation for $2m$-point total amplitudes are also satisfied. In section 6, we summarize this work. Useful diagrams and convention of notations are included in appendix.

\section{Preparation: Feynman rules and Berends-Giele recursion}
In this section, we review the Feynman rules and Berends-Giele recursion in non-linear sigma model which are
useful through this paper. Most of the notations follow the recent works \cite{Kampf:2012fn,Kampf:2013vha}.

\subsection{Feynman rules}

{~~~~~\emph {Lagrangian}}

The Lagrangian of $U(N)$ non-linear sigma model is given as
\bea
\mathcal{L}={F^2\over 4}\Tr (\partial_{\mu}U\partial^{\mu}U^{\dagger}),
\eea
where $F$ is a constant. As in \cite{Kampf:2012fn,Kampf:2013vha}, we use Cayley parametrization.
Under Cayley parametrization $U$ is defined as
\bea
U=1+2\Sl_{n=1}^{\infty}\left({1\over 2F}\phi\right)^n.~~~~\label{Cayley}
\eea
Here $\phi=\sqrt{2}\phi^at^a$ and $t^a$ are generators of $U(N)$ Lie algebra.

{\emph {Trace form of color decomposition}}

The total tree amplitudes can be given in terms of color-ordered amplitudes by trace form of color decomposition
\bea
M(1^{a_1},\dots,n^{a_n})=\Sl_{\sigma\in S_{n-1}}\Tr(T^{a_{1}}T^{a_{\sigma_2}}\dots T^{a_{\sigma_n}})A(1,\sigma).\label{Trace form}
\eea
Since the traces have cyclic symmetry, the color-ordered amplitudes also satisfy cyclic symmetry
\bea
A(1,2,\dots,n)=A(n,1,\dots,n-1).\label{Cyclic symmetry}
\eea

\emph{Feynman rules for color-ordered amplitudes}

Vertices in color-ordered Feynman rules under Cayley parametrization \eqref{Cayley} are
\bea
V_{2n+1}=0, V_{2n+2}=\left(-{1\over 2F^2}\right)^n\left(\Sl_{i=0}^np_{2i+1}\right)^2=\left(-{1\over 2F^2}\right)^n\left(\Sl_{i=0}^np_{2i+2}\right)^2,
\eea
where momentum conservation has been considered.

\subsection{Berends-Giele recursion}
In the Feynman rule given by the previous subsection, one can construct tree-level currents\footnote{In this paper, an $n$-point current is mentioned as the current with $n-1$ on-shell legs and one off-shell leg.} with one off-shell line through Berends-Giele recursion
\bea
&&J(2,...,n)\nn
&=&\frac{i}{P_{2,n}^2}\Sl_{m=4}^n\Sl_{1=j_0<j_1<\cdots<j_{m-1}=n}i V_{m}(p_1=-P_{2,n},P_{j_0+1,j_1},\cdots,P_{j_{m-2}+1,n})\times\prod\limits_{k=0}^{m-2} J(j_k+1,\cdots,j_{k+1}),\label{B-G}\nn
\eea
where $p_1=-P_{2,n}=-(p_2+p_3+\dots+p_n)$. The starting point of this recursion is $J(2)=J(3)=\dots=J(n)=1$.

 There is at least one odd-point vertex for current with odd-point lines(including the off-shell line) and the odd-point vertices are zero. As a result, we have
\bea
J(2,\dots,2m+1)=0,
\eea
for $(2m+1)$-point amplitudes.
The currents with even-points in general are nonzero and are built up by only odd numbers of even-point sub-currents. Since odd-point currents have to vanish,
in all following sections of this paper, we just need to discuss on the relations among even-point currents.

\section{Off-shell and on-shell $U(1)$ identity from Berends-Giele recursion}
In this section, we prove the $U(1)$ identity satisfied by even-point currents.
The identity is given as
\bea
\Sl_{\sigma\in OP(\{\alpha_1\}\bigcup\{\beta_1,\dots,\beta_{2m}\})}J(\{\sigma\})={1\over 2F^2}\Sl_{divisions\{\beta\}\rightarrow\{B_{1}\},\{B_{2}\}}J(\{B_{1}\})J(\{B_{2}\}),\label{off-shell-U(1)}
\eea
where, on the left hand side, we sum over all the possible permutations with keeping the relative orders in $\{\beta\}$ set and there is only one element $\alpha_1$ in $\{\alpha\}$ set. On the right hand side, we divide the ordered set $\{\beta_1,\dots,\beta_{2m}\}$ into two nonempty subsets. In each subset, there are odd number of $\beta$'s. For example, if there are six $\beta$'s, there are three possible divisions $\{B_1\}=\{\beta_1\}$, $\{B_{2}\}=\{\beta_2,\dots,\beta_6\}$; $\{B_1\}=\{\beta_1,\beta_2,\beta_3\}$, $\{B_{2}\}=\{\beta_4,\beta_5,\beta_6\}$ and $\{B_1\}=\{\beta_1,\dots,\beta_5\}$, $\{B_{2}\}=\{\beta_6\}$.

When we want to get the on-shell relations between amplitudes from the identity \eqref{off-shell-U(1)}, we should multiply both sides of \eqref{off-shell-U(1)}
by $p_1^2=(p_{\alpha_1}+p_{\beta_1}+\dots +p_{\beta_{2m}})^2$ and take the limit $p_1^2\rightarrow 0$. Since the right hand side are products of currents which are finite when $p_1^2$ goes to zero, after multiplied by $p_1^2$, the right hand side has to vanish under $p_1^2\rightarrow 0$. Then we arrive at on-shell $U(1)$-decoupling identity immediately
\bea
\Sl_{\sigma\in OP(\{\alpha_1\}\bigcup\{\beta_1,\dots,\beta_{2m}\})}A(1,\{\sigma\})=0.~~\label{on-shell-U(1)}
\eea

It is worth comparing the $U(1)$ identities in non-linear sigma model and in Yang-Mills theory. In Yang-Mills theory, $U(1)$-decoupling \cgg{identity}{identities} in both on-shell and off-shell cases have the same form. Thus, in both off-shell and on-shell cases, the \cgg{identity}{identities} can be understood as the decoupling of $U(1)$-gauge field. However, in non-linear sigma model, the $U(1)$ field can only decouple in the on-shell case. In off-shell case, at least for the choice of Cayley parametrization, we get sum of products of two sub-currents. In other words, only when taking the on-shell limit, the $U(1)$ field decouples.

Before proving the identity \eqref{off-shell-U(1)}, let us have a look at two examples.

\subsection{Four-point example}
In four-point case, the U(1)-identity is
\bea
J(\alpha_1,\beta_1,\beta_2)+J(\beta_1,\alpha_1,\beta_2)+J(\beta_1,\beta_2,\alpha_1)=\frac{1}{2F^2}J(\beta_1)J(\beta_2)=\frac{1}{2F^2}.
\eea
This is easy to prove by substituting the four-point vertex into the left hand side directly
\bea
&&J(\alpha_1,\beta_1,\beta_2)+J(\beta_1,\alpha_1,\beta_2)+J(\beta_1,\beta_2,\alpha_1)\nn
&=&-\frac{1}{2F^2}\frac{i}{p_1^2}i\left[(p_{1}+p_{\beta_1})^2+(p_1+p_{\alpha_1})^2+(p_1+p_{\beta_2})^2\right]\nn
&=&-\frac{1}{2F^2}\frac{i}{p_1^2}i\left[p_1^2+p_{\alpha_1}^2+p_{\beta_1}^2+p_{\beta_2}^2\right]\nn
&=&\frac{1}{2F^2}.
\eea
where $1$ is the off-shell line and we have used the on-shell conditions $p^2_{\alpha_1}=0$, $p^2_{\beta_1}=0$, $p^2_{\beta_2}=0$.

\subsection{Eight-point example}
Now let us skip the proof of six-point $U(1)$ identity and show how to use lower-point identity to prove eight-point $U(1)$ identity. The eight-point $U(1)$ identity is given as
\bea
&&\Sl_{\sigma\in OP(\{\alpha_1\}\bigcup\{\beta_1,\dots,\beta_6\})}J(\{\sigma\})\nn
&=&\frac{1}{2F^2}\left[J(\beta_1)J(\beta_2,\dots,\beta_6)+J(\beta_1,\beta_2,\beta_3)J(\beta_4,\beta_5,\beta_6)+J(\beta_1,\dots,\beta_5)J(\beta_6)\right]\nn
&=&\frac{1}{2F^2}\left[J(\beta_2,\dots,\beta_6)+J(\beta_1,\beta_2,\beta_3)J(\beta_4,\beta_5,\beta_6)+J(\beta_1,\dots,\beta_5)\right].\label{8pt-offshell-U(1)}
\eea
\begin{figure}
  \centering
 \includegraphics[width=0.6\textwidth]{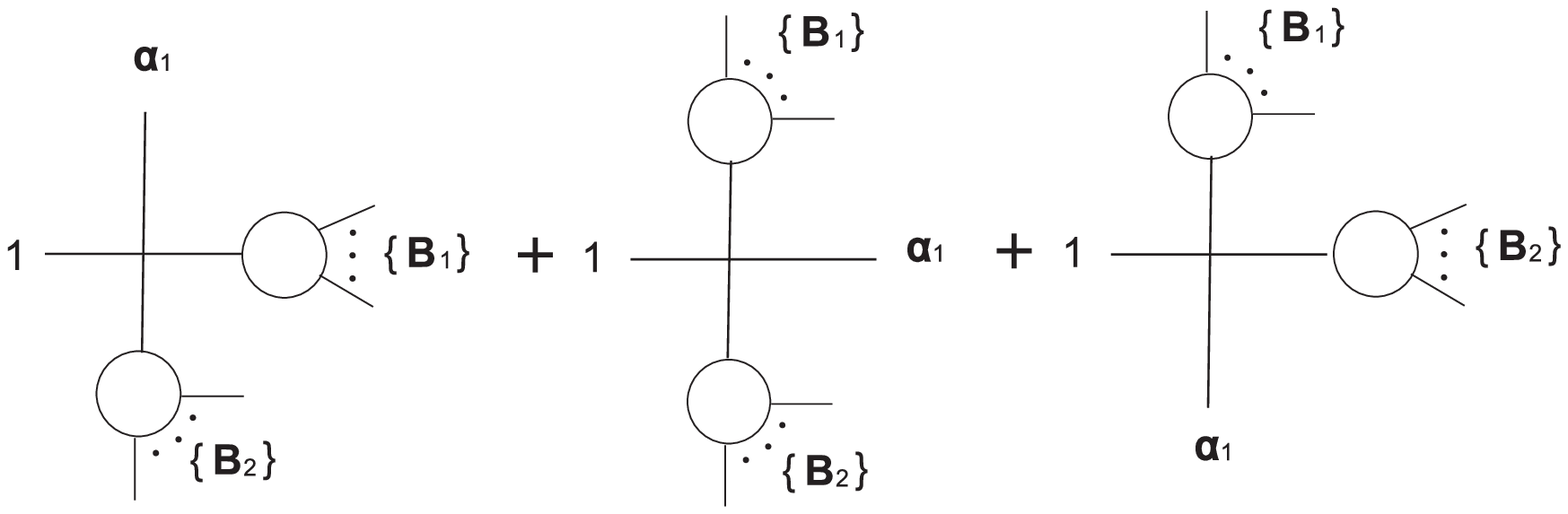}
 \caption{Sum of  diagrams with $\alpha_1$ connected with the off-shell leg directly via four-point vertex in $U(1)$ identity.} \label{u1_sub1}
\end{figure}
\begin{figure}
  \centering
 \includegraphics[width=0.4\textwidth]{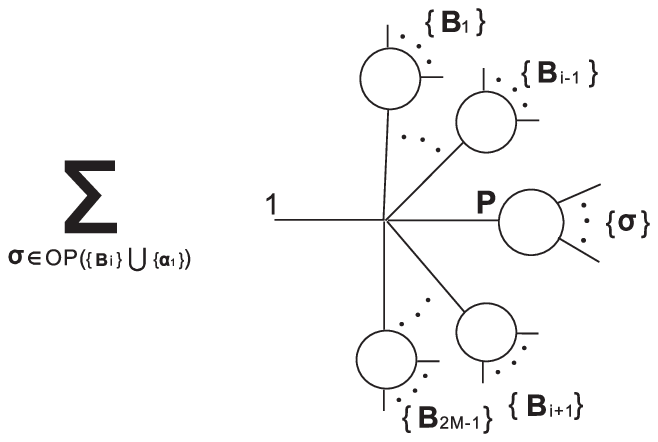}
 \caption{A diagram with lower-point substructure of $U(1)$ identity.} \label{u1_sub2}
\end{figure}
To prove this relation, we first show the explicit expression of Fig. \ref{u1_sub1} and Fig.  \ref{u1_sub2}.
\begin{itemize}
\item Fig. \ref{u1_sub1} can be expressed as
\bea
&&\text{Fig. \ref{u1_sub1}}\nn
&=&-\frac{1}{2F^2}\frac{i}{p_1^2}i\left[p_1^2+p_{\alpha_1}^2+p_{B_1}^2+p_{B_2}^2\right]J(\{B_1\})J(\{B_2\})\nn
&=&\frac{1}{2F^2}J(\{B_1\})J(\{B_2\})+\frac{1}{2F^2}{1\over p_1^2}\left[p_{B_1}^2J(\{B_1\})\right]J(\{B_2\})+\frac{1}{2F^2}J(\{B_1\}){1\over p_1^2}\left[p_{B_2}^2J(\{B_2\})\right]\nn
&=&\frac{1}{2F^2}J(\{B_1\})J(\{B_2\})\nn
&&+{1\over p_1^2}\Sl_{divisions\{B_1\}\rightarrow\{B_{1_1}\}\dots\{B_{1_{2i+1}}\}}\left(-\frac{1}{2F^2}\right)^{i+1}V_{2i+2}(-P_{B_{1_1},B_{1_{2i+1}}},P_{B_{1_1}},\dots P_{B_{1_i}})\nn
&&~~~~~~~~~~~~~~~~~~~~~~~~~~~~~~~~~~~~~~~~~~~~~~~~~~~~~~~~~~~~~\times J(\{B_2\})J(\{B_{1_1}\})\dots J(\{B_{1_{2i+1}}\})\nn
&&+{1\over p_1^2}\Sl_{divisions\{B_2\}\rightarrow\{B_{2_1}\}\dots\{B_{2_{2i+1}}\}}\left(-\frac{1}{2F^2}\right)^{i+1}V_{2i+2}(-P_{B_{2_1},B_{2_{2i+1}}},P_{B_{2_1}},\dots P_{B_{2_i}})\nn
&&~~~~~~~~~~~~~~~~~~~~~~~~~~~~~~~~~~~~~~~~~~~~~~~~~~~~~~~~~~~~~\times J(\{B_1\})J(\{B_{2_1}\})\dots J(\{B_{2_{2i+1}}\}),\label{U(1)-prop1}
\eea
where we have used the on-shell condition $p_{\alpha_1}^2=0$. $p_{B_i}$ denotes the sum of momenta of the on-shell lines in the set $\{B_i\}$.

\item Fig.  \ref{u1_sub2} can be expressed explicitly by using lower-point $U(1)$ identities
\bea
\text{Fig. \ref{u1_sub2}}&=&\Sl_{divisions\{B_i\}\rightarrow\{B_{i_1}\}\{B_{i_2}\}}\left(-1\over 2F^2\right)^{M}{1\over p_1^2}V(p_1,p_{B_1},\dots,p_{B_{i-1}},p_{B_i},p_{B_{i+1}},\dots,p_{B_{2M-1}})\nn
&&
~~~~~~~~~~~~~~\times J(\{B_1\})\dots J(\{B_{i-1}\})J(\{B_{i_1}\})J(\{B_{i_2}\})J(\{B_{i+1}\})\dots J(\{B_{2M-1}\}).\label{U(1)-prop2}
\eea
\end{itemize}

By Berends-Giele recursion, we can express the left hand side of eight-point $U(1)$ identity \eqref{8pt-offshell-U(1)} by sum of the diagrams in Fig. \ref{8pt_diagrams}.
We can always use \eqref{U(1)-prop2} to reduce sum of the terms with sub-currents containing both $\alpha_1$ and elements in $\{\beta\}$ into products of currents with only $\beta$ element. Thus the left hand side of \eqref{8pt-offshell-U(1)} can be expressed in terms of $J(\{B_1\})\dots J(\{B_{2M}\})$, where $\{B_1\}\dots\{B_{2M}\}$ is an nontrivial division of $\{\beta\}$. Each subset of this division must containing odd number of $\beta$ elements because the odd-point current must vanish.  We can classify the products of sub-currents into three categories according to different  number of sub-currents
\begin{itemize}
\item six sub-currents: $J(\beta_1)\dots J(\beta_6)$
\item four sub-currents: $J(\beta_1)J(\beta_2)J(\beta_3)J(\beta_4,\beta_5,\beta_6)$,
$J(\beta_1)J(\beta_2)J(\beta_3,\beta_4,\beta_5)J(\beta_6)$,

~~~~~~~~~~~~~~~~~~~~~~~~$J(\beta_1)J(\beta_2,\beta_3,\beta_4)J(\beta_5)J(\beta_6)$
and $J(\beta_1,\beta_2,\beta_3)J(\beta_4)J(\beta_5)J(\beta_6)$
\item two sub-currents: $J(\beta_1)J(\beta_2,\dots,\beta_6)$, $J(\beta_1,\beta_2,\beta_3)J(\beta_4,\beta_5,\beta_6)$ and $J(\beta_1,\dots,\beta_5)J(\beta_6)$.
\end{itemize}
Now let us discuss these contributions one by one.

{\bf(i)} Six sub-currents: $J(\beta_1)J(\beta_2)J(\beta_3)J(\beta_4)J(\beta_5)J(\beta_6)=1$.  There are three parts of contributions $\mathbb{A}$, $\mathbb{B}$ and $\mathbb{C}$ in this case.

$\mathbb{A}$ part is (A.1) in Fig. \ref{8pt_diagrams} and can be given as
 \bea
\mathbb{A}&=& i{i\over p_1^2}\left(-{1\over 2F^2}\right)^3 \bigl[(p_{\alpha_1}+p_{\beta_2}+p_{\beta_4}+p_{\beta_6})^2+(p_{\beta_1}+p_{\beta_2}+p_{\beta_4}+p_{\beta_6})^2+(p_{\beta_1}+p_{\alpha_1}+p_{\beta_4}+p_{\beta_6})^2\nn
&+&(p_{\beta_1}+p_{\beta_3}+p_{\beta_4}+p_{\beta_6})^2+(p_{\beta_1}+p_{\beta_3}+p_{\alpha_1}+p_{\beta_6})^2+(p_{\beta_1}+p_{\beta_3}+p_{\beta_5}+p_{\beta_6})^2\nn
&+&(p_{\beta_1}+p_{\beta_3}+p_{\beta_5}+p_{\alpha_1})^2\bigr].
 \eea

$\mathbb{B}$ part is sum of (B.5), (B.6), (B.7), (B.8) and (B.9) in  Fig. \ref{8pt_diagrams}. Using the property \eqref{U(1)-prop2}, this part can be given as
\bea
\mathbb{B}&=&\left({1\over 2F^2}\right)^3i{i\over p_1^2}\bigl[(p_{\alpha_1}+p_{\beta_1}+p_{\beta_2}+p_{\beta_4}+p_{\beta_6})^2+(p_{\beta_1}+p_{\beta_4}+p_{\beta_6})^2+(p_{\beta_1}+p_{\alpha_1}+p_{\beta_3}+p_{\beta_4}+p_{\beta_6})^2\nn
&+&(p_{\beta_1}+p_{\beta_3}+p_{\beta_6})^2+(p_{\beta_1}+p_{\beta_3}+p_{\alpha_1}+p_{\beta_5}+p_{\beta_6})^2\bigr].
\eea

$\mathbb{C}$ part gets contributions from the diagrams (C.1) and (C.3). Particularly, we apply the property \eqref{U(1)-prop1} to these two diagrams,
 then we find that the division $\{\beta_2,\beta_3,\beta_4,\beta_5,\beta_6\}\rightarrow\{\beta_2\},\{\beta_3\},\{\beta_4\},\{\beta_5\},\{\beta_6\}$ of (C.1) and
 the division $\{\beta_1,\beta_2,\beta_3,\beta_4,\beta_5\}\rightarrow\{\beta_1\},\{\beta_2\},\{\beta_3\},\{\beta_4\},\{\beta_5\}$ of (C.3) contribute to this case.
$\mathbb{C}$ can be expressed as
\bea
\mathbb{C}={i\over p_1^2}i\left({1\over 2F^2}\right)^3(p_{\beta_2}+p_{\beta_4}+p_{\beta_6})^2+{i\over p_1^2}i\left({1\over 2F^2}\right)^3(p_{\beta_1}+p_{\beta_3}+p_{\beta_5})^2.
\eea

Considering all three parts, we find that
\bea
\mathbb{A}+\mathbb{B}+\mathbb{C}={1\over p_1^2}\left({1\over 2F^2}\right)^3p_{\alpha_1}^2=0,
\eea
where we have used the on-shell condition of $\alpha_1$.

{\bf(ii)} Four sub-currents: There are four different products of sub-currents $J(\beta_1,\beta_2,\beta_3)J(\beta_4)J(\beta_5)J(\beta_6)$, $J(\beta_1)J(\beta_2,\beta_3,\beta_4)J(\beta_5)J(\beta_6)$,
$J(\beta_1)J(\beta_2)J(\beta_3,\beta_4,\beta_5)J(\beta_6)$ and $J(\beta_1)J(\beta_2)J(\beta_3)J(\beta_4,\beta_5,\beta_6)$.
 Now let us consider $J(\beta_1,\beta_2,\beta_3)J(\beta_4)J(\beta_5)J(\beta_6)$ as an example. The contributions of this case can also be classified
 into three parts $\mathbb{A}$, $\mathbb{B}$, $\mathbb{C}$.

 $\mathbb{A}$ part is given by (B.1) in  Fig. \ref{8pt_diagrams} and can be expressed explicitly
 \bea
 \mathbb{A}&=&i{i\over p_1^2}\left({1\over 2F^2}\right)^2\bigl[(p_{\alpha_1}+p_{\beta_4}+p_{\beta_6})^2+(p_{\beta_1}+p_{\beta_2}+p_{\beta_3}+p_{\beta_4}+p_{\beta_6})^2+(p_{\beta_1}+p_{\beta_2}+p_{\beta_3}+p_{\alpha_1}+p_{\beta_6})^2\nn
 &+&(p_{\beta_1}+p_{\beta_2}+p_{\beta_3}+p_{\beta_5}+p_{\beta_6})^2+(p_{\beta_1}+p_{\beta_2}+p_{\beta_3}+p_{\beta_5}+p_{\alpha_1})^2\bigr].
 \eea

$\mathbb{B}$ part get contributions from (C.4), (C.11) and (C.12) in  Fig. \ref{8pt_diagrams}. Particularly, we apply the property \eqref{U(1)-prop2}
to (C.4), (C.11) and (C.12). Then (C.11), (C.12) and the division $\{\beta_1,\beta_2,\beta_3,\beta_4,\beta_5\}\rightarrow \{\beta_1\},\{\beta_2\},\{\beta_3\},\{\beta_4\},\{\beta_5\}$ of (C.4) contribute to $\mathbb{B}$. Thus $\mathbb{B}$ can be given as
\bea
\mathbb{B}&=&-\left({1\over 2F^2}\right)^2i{i\over p_1^2}(p_{\alpha_1}+p_{\beta_1}+p_{\beta_2}+p_{\beta_3}+p_{\beta_4}+p_{\beta_6})^2-\left({1\over 2F^2}\right)^2i{i\over p_1^2}(p_{\beta_6}+p_{\beta_1}+p_{\beta_2}+p_{\beta_3})^2\nn
&&-\left({1\over 2F^2}\right)^2i{i\over p_1^2}(p_{\beta_1}+p_{\beta_2}+p_{\beta_3}+p_{\alpha_1}+p_{\beta_5}+p_{\beta_6})^2.
\eea

$\mathbb{C}$ part gets contributions from (C.2) and (C.3). Particularly, when applying \eqref{U(1)-prop1} to (C.2) and (C.3). The divisions $\{\beta_4,\beta_5,\beta_6\}\rightarrow\{\beta_4\},\{\beta_5\},\{\beta_6\}$ of (C.2) and $\{\beta_1,\beta_2,\beta_3,\beta_4,\beta_5\}\rightarrow\{\beta_1,\beta_2,\beta_3\},\{\beta_4\},\{\beta_5\}$ of (C.3) contribute to this case. Thus $\mathbb{C}$ part
is given as
\bea
\mathbb{C}=-i{i\over p_1^2}\left({1\over 2F^2}\right)^2\left[(p_{\beta_4}+p_{\beta_6})^2+(p_{\beta_1}+p_{\beta_2}+p_{\beta_3}+p_{\beta_5})^2\right].
\eea

Taking all three parts into account, we get
 \bea
 \mathbb{A}+\mathbb{B}+\mathbb{C}=0,
 \eea
 where we have used on-shell condition of $\alpha_1$. Following a similar way, we find that the other products of four sub-currents also cancel out.

{\bf(iii)} Two sub-currents: There are three non-vanishing products of sub-currents  $J(\beta_1)J(\beta_2,\dots,\beta_6)$, $J(\beta_1,\beta_2,\beta_3)J(\beta_4,\beta_5,\beta_6)$ and $J(\beta_1,\dots,\beta_5)J(\beta_6)$.
They can only get contributions from the three diagrams (C.1), (C.2) and (C.3).
Particularly, we apply the property \eqref{U(1)-prop1} to (C.1), (C.2) and (C.3). In this case, we need to keep the terms that of $(p_1^2)^0$ in these three diagrams. Then we get
\bea
\frac{1}{2F^2}\left[J(\beta_1)J(\beta_2,\dots,\beta_6)+J(\beta_1,\beta_2,\beta_3)J(\beta_4,\beta_5,\beta_6)+J(\beta_1,\dots,\beta_5)J(\beta_6)\right],
\eea
which is just the right hand side of the $U(1)$ identity for eight-point currents.

Therefore, after considering all the cases (i) (ii) and (iii), we get the $U(1)$ identity \eqref{8pt-offshell-U(1)} for eight-point currents.

\subsection{General proof}

Having shown the proof of the eight-point example, let us extend the proof to the general form of $U(1)$ identity. In general, one can always express the left hand side of \eqref{off-shell-U(1)} by lower-point sub-currents via Berends-Giele recursion \eqref{B-G}.   As in the eight-point examples, we can collect the diagrams with same off-shell momenta of sub-currents together. Then we can use the property \eqref{U(1)-prop2} to reduce the diagrams containing a substructure of $U(1)$ identity (as shown in Fig. \ref{u1_sub2}). After these reductions, the sub-currents containing both $\alpha_1$ and $\{\beta\}$ elements are reduced to products of sub-currents with only elements in $\{\beta\}$ set. Furthermore, we can apply \eqref{U(1)-prop1} to a four-point structure in Fig. \ref{u1_sub1}.
After these reductions, we should read out the coefficients of $J(\{B_1\})\dots J(\{B_{2M}\})$ for an arbitrary nontrivial division $\{\beta_1,\dots,\beta_{2m}\}\rightarrow\{B_1\}\dots\{B_{2M}\}$.

For $M>1$, as shown in the eight-point case, there are always three types of contributions Type-A, Type-B and Type-C in Fig. \ref{gen}. The notations in these diagrams are defined by Fig. \ref{convention3}.

For Type-A diagrams in Fig. \ref{gen}, we can always \cgg{used}{use} Feynman rules and momentum conservation to avoid the appearance of the momentum of the off-shell leg $1$ and express the coefficient of $J(\{B_1\})\dots J(\{B_{2M}\})$  by the on-shell momenta.

For Type-B diagrams in Fig. \ref{gen}, as have mentioned, we should substitute \eqref{U(1)-prop2} into these diagrams to reduce them and keep the right divisions
that can produce $J(\{B_1\})\dots J(\{B_{2M}\})$. For example, we should keep the division $\{B_1,B_2\}\rightarrow\{B_1\},\{B_2\}$ in the first diagram and keep the division $\{B_2,B_3\}\rightarrow\{B_2\},\{B_3\}$ in the second diagram, and so on. For convenience, we also express the vertices in Type-B diagrams by the on-shell momenta via momentum conservation.

For Type-C diagrams in Fig. \ref{gen}, we should apply \eqref{U(1)-prop1}. For the first diagram of Type-C, we should keep the division $\{B_2,\dots,B_{2M}\}\rightarrow\{B_2\}\dots\{B_{2M}\}$ while, for the second diagram we should keep the division $\{B_1,\dots,B_{2M-1}\}\rightarrow\{B_1\}\dots\{B_{2M-1}\}$.

\begin{table}
{\tiny\centering\begin{tabular}{|c|c|c|c|c|c|c|c|c|}
  \hline
    & $p_{\alpha_1}^2$ & $p_{B_{2i+1}}^2$ & $p_{B_{2i}}^2$ & $s_{\alpha_1B_{2i+1}}$ & $s_{\alpha_1B_{2i}}$ & $s_{B_{2i+1}B_{2j+1}}$ & $s_{B_{2i}B_{2j}}$ & $s_{B_{2i+1}B_{2j}}$ \\ \hline
  Type-A & $(M+1)$ & $2(M-i)$ & $2i$       &$M-i$   & $i$ & $\Bigl\{
                                                              \begin{array}{c}
                                                                2(M-j)~(i<j) \\
                                                                0~~(\text{Otherwise})\\                                                                                                              \end{array}$
   &$\Bigl\{                                                  \begin{array}{c}
                                                                2i~(i<j) \\
                                                                0~(\text{Otherwise})\\                                                                                                              \end{array}$  &$\Bigl\{\begin{array}{c}
                                                                2(j-i)-1~(i<j) \\
                                                                0~(\text{Otherwise})\\                                                                                                              \end{array}$ \\ \hline
  Type-B & $-M$    & $-2(M-i)+1$&$-2i+1$   &$-M+i$  & $-i$ &$\Bigl\{\begin{array}{c}
                                                                -2(M-j)+1~(i<j) \\
                                                                0~~(\text{Otherwise})\\                                                                                                              \end{array}$  &$\Bigl\{\begin{array}{c}
                                                                -2i+1~(i<j) \\
                                                                0~~(\text{Otherwise})\\                                                                                                              \end{array}$ &$\Bigl\{\begin{array}{c}
                                                                -2(j-i)+1~(i<j) \\
                                                                0~~(\text{Otherwise})\\                                                                                                              \end{array}$   \\ \hline
  Type-C & 0       & -1        &   -1      &$0$     & $0$ &$\Bigl\{\begin{array}{c}
                                                                -1~(i<j) \\
                                                                0~~(\text{Otherwise})\\                                                                                                              \end{array}$  & $\Bigl\{\begin{array}{c}
                                                                -1~(i<j) \\
                                                                0~~(\text{Otherwise})\\                                                                                                              \end{array}$& 0 \\
  \hline
\end{tabular}}\caption{Coefficients of $J(\{B_1\})\dots J(\{B_{2M}\})$ in $U(1)$ identity. Here $s_{\alpha_1B_u}$ denotes $2p_{\alpha_1}\cdot\left(\Sl_{\beta_p\in\{B_u\}} p_{\beta_p}\right)$,$s_{B_uB_v}$ denotes $2\left(\Sl_{\beta_p\in\{B_u\}} p_{\beta_p}\right)\cdot\left(\Sl_{\beta_q\in\{B_v\}} p_{\beta_q}\right)$, $u$, $v$ can be $2i$ or $2i+1$. For $B_{2i+1}$, $i$ runs from $0$ to $M-1$, while for $B_{2i}$, $i$ runs from $1$ to $M$.}\label{t1}\end{table}

Then we can collect all the coefficients in the three types in Table \ref{t1}.
In Table \ref{t1}, we have left a total factor ${i\over p_1^2}i\left(-{1\over 2F^2}\right)^M$ apart. Thus, the total coefficient of $J(\{B_1\})J(\{B_2\})\dots J(\{B_{2M}\})$ is ${i\over p_1^2}i\left(-{1\over 2F^2}\right)^M p_{\alpha_1}^2$. Since $p_{\alpha_1}^2=0$, the $J(\{B_1\})J(\{B_2\})\dots J(\{B_{2M}\})$ must vanish.

For $M=1$, there are only two sub-currents in the products. In this case, we only need to consider the terms with $(p_1^2)^0$ in the diagrams of the form in Fig. \ref{u1_sub1}.  We should sum over all the possible $\{B_1\}$ and $\{B_2\}$ and get
\bea
p_1^2{i\over p_1^2}i\left(-{1\over 2F^2}\right)\Sl_{divisions\{\beta\}\rightarrow\{B_{1}\},\{B_{2}\}}J(\{B_{1}\})J(\{B_{2}\}),
\eea
which is just the right hand side of the off-shell $U(1)$ identity \eqref{off-shell-U(1)}.

\section{Off-shell and on-shell fundamental BCJ relation from Berends-Giele recursion}

Having proven the $U(1)$ identity, let us consider a more nontrivial relation-fundamental BCJ relation- in non-linear sigma model. Since the odd-point currents and amplitudes must vanish, we only need to consider
the relations for even-point currents and amplitudes. Being different from $U(1)$ identity, fundamental BCJ relation has non-trivial coefficients
accompanying with the currents or amplitudes. The general formula of off-shell fundamental BCJ relation is given as
\bea
&&\Sl_{\sigma\in OP(\{\alpha_1\}\bigcup\{\beta_1,\dots,\beta_{2m-1}\})}\Sl_{\xi_{\sigma_i}<\xi_{\alpha_1}}s_{\alpha_1\sigma_i}J(\{\sigma\},\beta_{2m})\nn
&=&-{1\over 2F^2}\Sl_{divisions\{\beta\}\rightarrow\{B_{1}\},\{B_{2}\}}\left[\Sl_{\beta_i\in\{B_{2}\}}s_{\alpha_1\beta_i}J(\{B_{1}\})J(\{B_{2}\})\right],\label{off-shell-BCJ}
\eea
where we use $\xi_i$ to denote the position of the leg $i$ in permutation $\sigma$, we define $\xi_1=0$, thus we always have a $s_{\alpha_11}$ in the coefficients for each currents on the left hand side. On the right hand side, we sum over all the possible divisions of the ordered set $\{\beta\}$ into two sub-ordered sets $\{B_{1}\}$ and $\{B_{2}\}$. Since $J(\{B_{1}\})$ or $J(\{B_{2}\})$ must vanish when $\{B_1\}$ or $\{B_{2}\}$ have even number, the divisions that survive are those with both odd number of elements in $\{B_{1}\}$ and $\{B_{2}\}$. Since the right hand side is finite under $p_1^2\rightarrow 0$,
after multiplying $p_1^2$ and taking the on-shell limit $p_1^2\rightarrow 0$ we get the on-shell relation for amplitudes
\bea
\Sl_{\sigma\in OP(\{\alpha_1\}\bigcup\{\beta_1,\dots,\beta_{2m-1}\})}\Sl_{\xi_{\sigma_i}<\xi_{\alpha_1}}s_{\alpha_1\sigma_i}A(1,\{\sigma\},\beta_{2m})=0.\label{on-shell-BCJ}
\eea

The left hand side of fundamental BCJ relation can be understood as following. We move one external leg $\alpha_1$ from the position next to the leg $1$ to the position in front of the leg $\beta_{2m}$. For each position, we can write down a corresponding current(or amplitude) accompanied by a kinematic factor $\Sl_{\xi_{\sigma_i}<\xi_{\alpha_1}}s_{\alpha_1\sigma_i}$. Then we sum over all the currents with coefficients.

Before giving the general proof of the relation \eqref{off-shell-BCJ}, let us have a look at two examples.

\subsection{Four-point example}
The simplest example is the four-point fundamental BCJ relation
\bea
&&s_{\alpha_11}J(\alpha_1,\beta_1,\beta_2)+(s_{\alpha_11}+s_{\alpha_1\beta_1})J(\beta_1,\alpha_1,\beta_2)=-\left(1\over 2F^2\right)s_{\alpha_1\beta_2}J(\beta_1)J(\beta_2).~~\label{4pt-offshell-BCJ}
\eea
To see this, we write the currents on the left hand side of BCJ relation \eqref{4pt-offshell-BCJ} explicitly via Feynman rules
\bea
&&s_{\alpha_11}J(\alpha_1,\beta_1,\beta_2)+(s_{\alpha_11}+s_{\alpha_1\beta_1})J(\beta_1,\alpha_1,\beta_2)\nn
&=&-\left(1\over 2F^2\right)i\frac{i}{p_1^2}\left[s_{\alpha_11}(p_{\alpha_1}+p_{\beta_2})^2+(s_{\alpha_11}+s_{\alpha_1\beta_1})(p_{\beta_1}+p_{\beta_2})^2\right]J(\beta_1)J(\beta_2)\nn
&=&-\left(1\over 2F^2\right)s_{\alpha_1\beta_2}J(\beta_1)J(\beta_2)\nn
&=&-\left(1\over 2F^2\right)s_{\alpha_1\beta_2},
\eea
where we have used momentum conservation and on-shell conditions of $\alpha_1$, $\beta_1$ and $\beta_2$. Thus we have proved the fundamental BCJ relation \eqref{4pt-offshell-BCJ} at four-point.

\subsection{Eight-point example}

The four-point example in above subsection just provides a starting point of  an inductive proof. In this subsection, we skip the proof of fundamental BCJ relation at six-point and assume that the relation \eqref{off-shell-BCJ} is satisfied at both four- and six- points. We will show how to prove the eight-point relation recursively.

Fundamental BCJ relation for eight-point currents is given as
\bea
&&\Sl_{\sigma\in OP(\{\alpha_1\}\bigcup\{\beta_1,\dots,\beta_{5}\})}\Sl_{\xi_{\sigma_i}<\xi_{\alpha_1}}s_{\alpha_1\sigma_i}J(\{\sigma\},\beta_{6})\nn
&=&-{1\over 2F^2}\Sl_{divisions\{\beta_1,\dots,\beta_6\}\rightarrow\{B_{1}\},\{B_{2}\}}\left[\Sl_{\beta_i\in\{B_{2}\}}s_{\alpha_1\beta_i}J(\{B_{1}\})J(\{B_{2}\})\right],~~\label{8pt-offshell-BCJ}
\eea
where, on the right hand side, we sum over three nonzero divisions $\{\beta_1,\dots,\beta_6\}\rightarrow\{\beta_1\}\{\beta_2,\beta_3,\beta_4,\beta_5,\beta_6\}$, $\{\beta_1,\dots,\beta_6\}\rightarrow\{\beta_1,\beta_2,\beta_3\}\{\beta_4,\beta_5,\beta_6\}$ and $\{\beta_1,\dots,\beta_6\}\rightarrow\{\beta_1,\beta_2,\beta_3,\beta_4,\beta_5\}\{\beta_6\}$.
\begin{figure}[!h]
  \centering
 \includegraphics[width=0.6\textwidth]{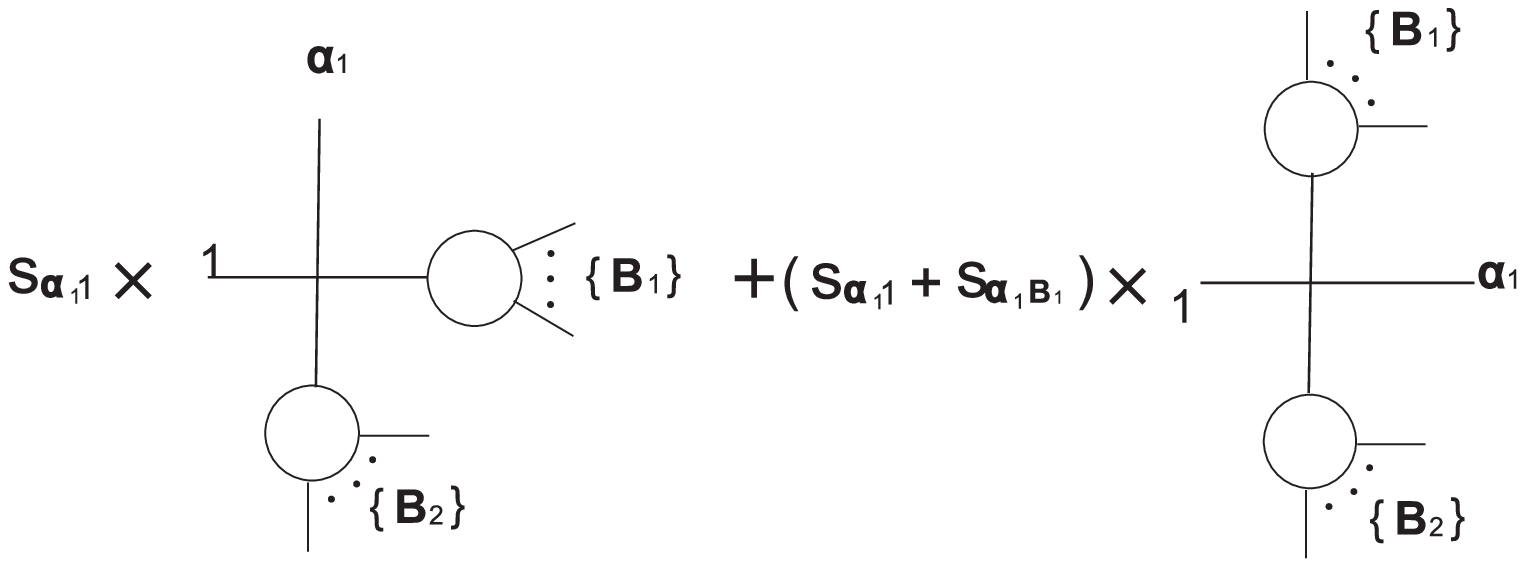}
 \caption{Sum of  diagrams with $\alpha_1$ connected with the off-shell leg directly via four-point vertex in BCJ relation.} \label{bcj_sub1}
\end{figure}
\begin{figure}[!h]
  \centering
 \includegraphics[width=0.9\textwidth]{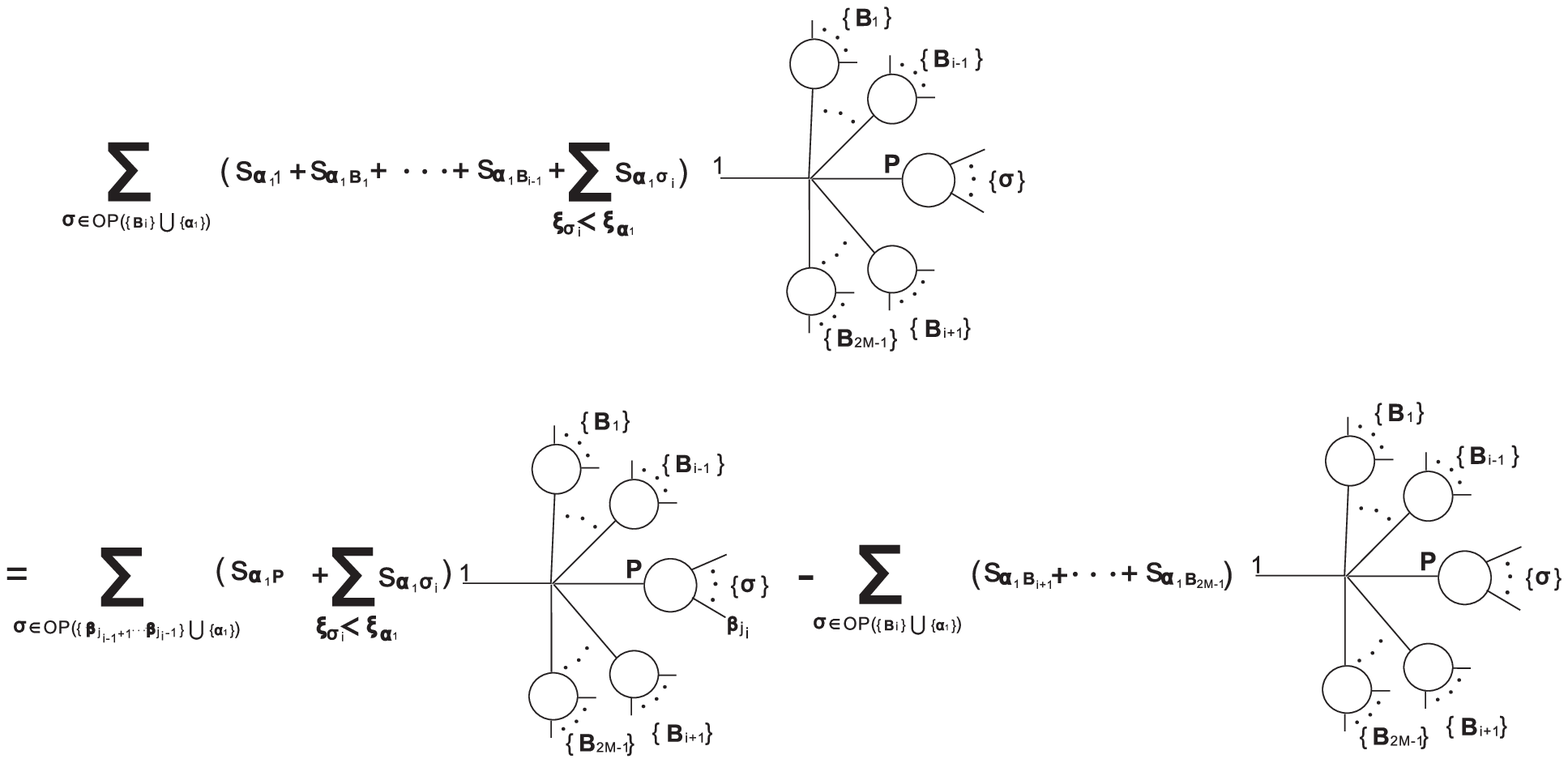}
 \caption{A diagram with lower-point substructure of BCJ relation.} \label{bcj_sub2}
\end{figure}

To prove this relation, we first show the explicit expressions of Fig. \ref{bcj_sub1} and Fig. \ref{bcj_sub2}:
\begin{itemize}
\item
We first consider the sum of the two diagrams in Fig. \ref{bcj_sub1}.
If we divide the ordered set $\{\beta_1,\dots,\beta_{2m}\}$ into two ordered subsets $\{B_1\}$ and $\{B_2\}$, then Fig. \ref{bcj_sub1} is given as
\bea
&&\text{Fig. \ref{bcj_sub1}}\nn
&=&{1\over 2F^2}{1\over p_1^2}\left[s_{\alpha_11}(p_{\alpha_1}+p_{B_2})^2+(s_{\alpha_11}+s_{\alpha_1B_1})(p_{B_1}+p_{B_2})^2\right]J(\{B_1\})J(\{B_2\})\nn
&=&{1\over 2F^2}{1\over p_1^2}(s_{\alpha_11}p_{B_2}^2-s_{\alpha_1B_2}p_1^2)J(\{B_1\})J(\{B_2\})\nn
&=&-{1\over 2F^2}{1\over p_1^2}s_{\alpha_1B_2}p_1^2J(\{B_1\})J(\{B_2\})\nn
&&+{1\over 2F^2}{1\over p_1^2}s_{\alpha_11}\Sl_{divisions\{B_2\}\rightarrow\{B_{2_1}\}\dots\{B_{2_{2i+1}}\}}\left(-\frac{1}{2F^2}\right)^iV_{2i+2}(-P_{B_{2_1},B_{2_{2i+1}}},P_{B_{2_1}},\dots P_{B_{2_{2i+1}}})\nn
&&~~~~~~~~~~~~~~~~~~~~~~~~~~~~~~~~~~~~~~~~~~~~~~~~~~~~~~~~~~~~~~~~~~\times J(\{B_1\})J(\{B_{2_1}\})\dots J(\{B_{2_{2i+1}}\}).\label{BCJ-prop1}
\eea

\item Now let us consider Fig. \ref{bcj_sub2}. The left hand side of Fig. \ref{bcj_sub2} can be reexpressed by the right hand side of Fig. \ref{bcj_sub2} by considering momentum conservation and on-shell condition of $\alpha_1$. Since the first and second terms of the right hand side of Fig. \ref{bcj_sub2} have substructures of fundamental BCJ relation and $U(1)$ identity respectively, we can further reduce them by lower-point relations. Then we have
    \bea
    \text{Fig. \ref{bcj_sub2}}&=&{1\over 2F^2}\Sl_{divisions\{B_i\}\rightarrow\{B_{i_1}\}\{B_{i_2}\}}\left(-1\over 2F^2\right)^{M-1}{1\over p_1^2}(s_{\alpha_1B_{i_2}}+s_{\alpha_1B_{i+1}}+\dots+s_{\alpha_1B_{2M-1}})\nn
    &&~~~~\times V(p_1,p_{B_1},\dots,p_{B_{i-1}},p_{B_i},p_{B_{i+1}},\dots,p_{B_{2M-1}})\nn
    &&~~~~\times J(\{B_1\})\dots J(\{B_{i-1}\})J(\{B_{i_1}\})J(\{B_{i_2}\})J(\{B_{i+1}\})\dots J(\{B_{2M-1}\}).\label{BCJ-prop2}
    \eea
    A special case is $i=2M-1$. In this case, $\alpha_i$ cannot be moved to the position next to the last element of $\{B_{2M-1}\}$. This case can also be included in Fig. \ref{bcj_sub2} by considering momentum conservation and on-shell condition of $\alpha_1$. Thus the property \eqref{BCJ-prop2} also holds.
\end{itemize}
With the above two properties, one can prove the eight-point fundamental BCJ relation \eqref{8pt-offshell-BCJ}. We can write the left hand side of
eight-point fundamental BCJ relation by lower-point currents via Berends-Giele recursion \eqref{B-G}. The left hand side of \eqref{8pt-offshell-BCJ} is given as sum of the diagrams in Fig. \ref{8pt_diagrams}. For the diagrams in Fig. \ref{8pt_diagrams}, we can apply \eqref{BCJ-prop2} to (B.5)-(B.9), (C.4)-(C.12) and apply \eqref{BCJ-prop1} to (C.1), (C.2), (C.3). It is easy to see that the left hand side of eight-point fundamental BCJ relation can be expressed in terms of products of currents of the form $J(\{B_1\})J(\{B_2\})\dots J(\{B_{2M}\})$ after considering the property \eqref{BCJ-prop2} and $J(\alpha_1)=1$, where $\{B_1\}\dots \{B_{2M}\}$ are non-vanishing divisions of the ordered set $\{\beta_1,\dots,\beta_6\}$. Then we can read off the coefficients for each division and prove the relation.

The divisions can be classified in following cases
\begin{itemize}
\item six sub-currents: $J(\beta_1)\dots J(\beta_6)$
\item four sub-currents: $J(\beta_1)J(\beta_2)J(\beta_3)J(\beta_4,\beta_5,\beta_6)$, $J(\beta_1)J(\beta_2)J(\beta_3,\beta_4,\beta_5)J(\beta_6)$,

~~~~~~~~~~~~~~~~~~~~~~~~$J(\beta_1)J(\beta_2,\beta_3,\beta_4)J(\beta_5)J(\beta_6)$ and $J(\beta_1,\beta_2,\beta_3)J(\beta_4)J(\beta_5)J(\beta_6)$
\item two sub-currents: $J(\beta_1)J(\beta_2,\dots,\beta_6)$, $J(\beta_1,\beta_2,\beta_3)J(\beta_4,\beta_5,\beta_6)$ and $J(\beta_1,\dots,\beta_5)J(\beta_6)$.
\end{itemize}
We can calculate the coefficients for these divisions one by one:

{\bf(i)} Six sub-currents: $J(\beta_1)J(\beta_2)J(\beta_3)J(\beta_4)J(\beta_5)J(\beta_6)=1$. This case get contributions from three parts $\mathbb{A}$, $\mathbb{B}$ and $\mathbb{C}$.

$\mathbb{A}$ part is (A.1) in Fig. \ref{8pt_diagrams} and can be given as
 \bea
\mathbb{A}&=& i{i\over p_1^2}\left(-{1\over 2F^2}\right)^3 \Bigl[s_{\alpha_11}(p_{\alpha_1}+p_{\beta_2}+p_{\beta_4}+p_{\beta_6})^2+(s_{\alpha_11}+s_{\alpha_1\beta_1})(p_{\beta_1}+p_{\beta_2}+p_{\beta_4}+p_{\beta_6})^2\nn
&&+(s_{\alpha_11}+s_{\alpha_1\beta_1}+s_{\alpha_1\beta_2})(p_{\beta_1}+p_{\alpha_1}+p_{\beta_4}+p_{\beta_6})^2\nn
&&+(s_{\alpha_11}+s_{\alpha_1\beta_1}+s_{\alpha_1\beta_2}+s_{\alpha_1\beta_3})(p_{\beta_1}+p_{\beta_3}+p_{\beta_4}+p_{\beta_6})^2\nn
&&+(s_{\alpha_11}+s_{\alpha_1\beta_1}+s_{\alpha_1\beta_2}+s_{\alpha_1\beta_3}+s_{\alpha_1\beta_4})(p_{\beta_1}+p_{\beta_3}+p_{\alpha_1}+p_{\beta_6})^2\nn
&&+(s_{\alpha_11}+s_{\alpha_1\beta_1}+s_{\alpha_1\beta_2}+s_{\alpha_1\beta_3}+s_{\alpha_1\beta_4}+s_{\alpha_1\beta_5})(p_{\beta_1}+p_{\beta_3}+p_{\beta_5}+p_{\beta_6})^2\Bigr].
 \eea
 $\mathbb{B}$ part is the sum of (B.5), (B.6), (B.7), (B.8) and (B.9) in Fig. \ref{8pt_diagrams}. Using the property \eqref{BCJ-prop2}, we get
\bea
\mathbb{B}&=&\left({1\over 2F^2}\right)^3i{i\over p_1^2}\Bigl[-(s_{\alpha_1\beta_2}+s_{\alpha_1\beta_3}+s_{\alpha_1\beta_4}+s_{\alpha_1\beta_5}+s_{\alpha_1\beta_6})(p_{\alpha_1}+p_{\beta_1}+p_{\beta_2}+p_{\beta_4}+p_{\beta_6})^2\nn
&&-(s_{\alpha_1\beta_3}+s_{\alpha_1\beta_4}+s_{\alpha_1\beta_5}+s_{\alpha_1\beta_6})(p_{\beta_1}+p_{\beta_4}+p_{\beta_6})^2\nn
&&-(s_{\alpha_1\beta_4}+s_{\alpha_1\beta_5}+s_{\alpha_1\beta_6})(p_{\beta_1}+p_{\alpha_1}+p_{\beta_3}+p_{\beta_4}+p_{\beta_6})^2\nn
&&-(s_{\alpha_1\beta_5}+s_{\alpha_1\beta_6})(p_{\beta_1}+p_{\beta_3}+p_{\beta_6})^2-s_{\alpha_1\beta_6}(p_{\beta_1}+p_{\beta_3}+p_{\alpha_1}+p_{\beta_5}+p_{\beta_6})^2\Bigr].
\eea
 $\mathbb{C}$ part is the division $\{\beta_2,\beta_3,\beta_4,\beta_5,\beta_6\}\rightarrow\{\beta_2\},\{\beta_3\},\{\beta_4\},\{\beta_5\},\{\beta_6\}$ of (C.1). Particularly, this part is given as
\bea
\mathbb{C}&=&-{i\over p_1^2}i\left({1\over 2F^2}\right)^3(s_{\alpha_1\beta_1}+s_{\alpha_1\beta_2}+s_{\alpha_1\beta_3}+s_{\alpha_1\beta_4}+s_{\alpha_1\beta_5}+s_{\alpha_1\beta_6})(p_{\beta_2}+p_{\beta_4}+p_{\beta_6})^2.\nn
\eea
Considering momentum conservation and on-shell condition $p_{\alpha_1}^2=0$, we can see $\mathbb{A}+\mathbb{B}+\mathbb{C}=0$.

{\bf(ii)} Four sub-currents: There are four different products of sub-currents $J(\beta_1,\beta_2,\beta_3)J(\beta_4)J(\beta_5)J(\beta_6)$, $J(\beta_1)J(\beta_2,\beta_3,\beta_4)J(\beta_5)J(\beta_6)$,
$J(\beta_1)J(\beta_2)J(\beta_3,\beta_4,\beta_5)J(\beta_6)$ and $J(\beta_1)J(\beta_2)J(\beta_3)J(\beta_4,\beta_5,\beta_6)$. Let us take $J(\beta_1,\beta_2,\beta_3)J(\beta_4)J(\beta_5)J(\beta_6)$ as an example.
$J(\beta_1,\beta_2,\beta_3)J(\beta_4)J(\beta_5)J(\beta_6)$ gets contributions from three parts $\mathbb{A}$, $\mathbb{B}$ and $\mathbb{C}$.

$\mathbb{A}$ part is the contribution of (B.1) in Fig. \ref{8pt_diagrams} and given as
\bea
\mathbb{A}&=&i{i\over p_1^2}\left({1\over 2F^2}\right)^2\Bigl[s_{\alpha_11}(p_{\alpha_1}+p_{\beta_4}+p_{\beta_6})^2+(s_{\alpha_11}+s_{\alpha_1\beta_1}+s_{\alpha_1\beta_2}+s_{\alpha_1\beta_3})(p_{\beta_1}+p_{\beta_2}+p_{\beta_3}+p_{\beta_4}+p_{\beta_6})^2\nn
&&+(s_{\alpha_11}+s_{\alpha_1\beta_1}+s_{\alpha_1\beta_2}+s_{\alpha_1\beta_3}+s_{\alpha_1\beta_4})(p_{\beta_1}+p_{\beta_2}+p_{\beta_3}+p_{\alpha_1}+p_{\beta_6})^2\nn
 &&+(s_{\alpha_11}+s_{\alpha_1\beta_1}+s_{\alpha_1\beta_2}+s_{\alpha_1\beta_3}+s_{\alpha_1\beta_4}+s_{\alpha_1\beta_5})(p_{\beta_1}+p_{\beta_2}+p_{\beta_3}+p_{\beta_5}+p_{\beta_6})^2\Bigr].
\eea

$\mathbb{B}$ is the sum of the (C.11), (C.12) in Fig. \ref{8pt_diagrams} and the division $\{\beta_1,\beta_2,\beta_3,\beta_4\}\rightarrow\{\beta_1,\beta_2,\beta_3\},\{\beta_4\}$ of (C.4) in Fig. \ref{8pt_diagrams}.
Particularly, we have
\bea
\mathbb{B}&=&-\left({1\over 2F^2}\right)^2i{i\over p_1^2}\Bigl[-(s_{\alpha_1\beta_5}+s_{\alpha_1\beta_6})(p_{\beta_6}+p_{\beta_1}+p_{\beta_2}+p_{\beta_3})^2-s_{\alpha_1\beta_6}(p_{\beta_1}+p_{\beta_2}+p_{\beta_3}+p_{\alpha_1}+p_{\beta_5}+p_{\beta_6})^2\nn
&&-(s_{\alpha_1\beta_4}+s_{\alpha_1\beta_5}+s_{\alpha_1\beta_6})(p_{\alpha_1}+p_{\beta_1}+p_{\beta_2}+p_{\beta_3}+p_{\beta_4}+p_{\beta_6})^2\Bigr].
\eea

$\mathbb{C}$ part gets contribution of division $\{\beta_4,\beta_5,\beta_6\}\rightarrow\{\beta_4\},\{\beta_5\},\{\beta_6\}$ of (C.2). This part is given as
\bea
\mathbb{C}&=&-i{i\over p_1^2}\left({1\over 2F^2}\right)^2\Bigl[-(s_{\alpha_1\beta_1}+s_{\alpha_1\beta_2}+s_{\alpha_1\beta_3}+s_{\alpha_1\beta_4}+s_{\alpha_1\beta_5}+s_{\alpha_1\beta_6})(p_{\beta_4}+p_{\beta_6})^2\Bigr].
\eea
After some calculations and considering momentum conservation and  on-shell conditions of the on-shell external lines, we get $\mathbb{A}+\mathbb{B}+\mathbb{C}=0$.
Following similar calculations, we find that coefficients for the other products of four-currents also vanish.

{\bf(iii)} Two sub-currents

In this case, only the terms that of  $(p_1^2)^0$ in (C.1), (C.2) and (C.3) contribute and the sum of these contributions is given as
\bea
&&\frac{1}{2F^2}\Bigl[-(s_{\alpha_1\beta_2}+s_{\alpha_1\beta_3}+s_{\alpha_1\beta_4}+s_{\alpha_1\beta_5}+s_{\alpha_1\beta_6})J(\beta_2,\dots,\beta_6)\nn
&-&(s_{\alpha_1\beta_4}+s_{\alpha_1\beta_5}+s_{\alpha_1\beta_6})J(\beta_1,\beta_2,\beta_3)J(\beta_4,\beta_5,\beta_6)\nn
&-&s_{\alpha_1\beta_6}J(\beta_1,\dots,\beta_5)\Bigr].
\eea
After considering all the cases (i), (ii) and (iii), we find that only the productions of two sub-currents are left and this part is just the right hand side  of eight-point fundamental BCJ relation.

\subsection{General proof}
Now let us consider the general proof of fundamental BCJ relation \eqref{off-shell-BCJ}. As shown in the eight-point example, we can always express the left hand side of the relation \eqref{off-shell-BCJ} by Berends-Giele recursion \eqref{B-G} and collect the diagrams with same off-shell momenta of sub-currents(e.g., for eight point case the diagrams are given by Fig. \ref{8pt_diagrams}). After applying \eqref{BCJ-prop1} and \eqref{BCJ-prop2}, the left hand side of \eqref{off-shell-BCJ} can be written in terms of $J(\{B_1\})\dots J(\{B_{2M}\})$, where $\{B_1\}\dots \{B_{2M}\}$ are nontrivial divisions\footnote{Since the odd-point current must vanish, the number of elements in each subset must be odd so that the product is nonzero.} of the ordered set $\{\beta\}$. To prove
the relation  \eqref{off-shell-BCJ}, we should read off the coefficient for each division. Then we should show that the coefficients must vanish for divisions with $M>1$ and must give the right hand side of \eqref{off-shell-BCJ} for divisions with $M=1$.

For given $M$ $(M>1)$, the diagrams contribute to $J(\{B_1\})\dots J(\{B_{2M}\})$ can be classified into three types (this is similar with the eight-point example) Type-A, Type-B and the first diagram of Type-C in Fig. \ref{gen}. The notations in these diagrams are defined by Fig. \ref{convention4}.

For Type-A diagram in Fig. \ref{gen} we can
use  momentum conservation and on-shell condition of $\alpha_1$ to rewrite the coefficient in each term into a form independent of momentum of the off-shell line  $1$. For example, if we consider the diagram with $\alpha_1$ between $\{B_i\}$ and $\{B_{i+1}\}$, the coefficient is rewritten as
\bea
s_{\alpha_11}+s_{\alpha_1B_1}+\dots+s_{\alpha_1B_i}=-(s_{\alpha_1B_{i+1}}+\dots+s_{\alpha_1B_{2M}}).
\eea
The vertex is also written in the form independent of the momentum of off-shell leg.

For Type-B diagrams in Fig. \ref{gen}, we should write down the expression of each diagram by \eqref{BCJ-prop2} and pick out the appropriate division such that we can get $\{B_1\}\dots \{B_{2M}\}$. For example, for the first diagram in Type-B in Fig. \ref{gen}, we should keep the division $\{B_1, B_2\}\rightarrow\{B_1\}\{B_2\}$ , for the second diagram we should keep the division $\{B_2, B_3\}\rightarrow\{B_2\}\{B_3\}$ and so on. We also write the coefficients and vertices as forms independent of the momentum of the off-shell leg $1$ via momentum conservation and on-shell condition of $\alpha_1$.

For Type-C diagrams in Fig. \ref{gen}, we should write down the expression of each diagram by \eqref{BCJ-prop1} and keep the divisions such that we can get $\{B_1\}\dots \{B_{2M}\}$. Only the first diagram of Type-C contributes. We should keep the division $\{B_2,\dots,B_{2M}\}\rightarrow\{B_2\}\dots\{B_{2M}\}$ of the first diagram of Type-C. We also use momentum conservation to rewrite $s_{\alpha_11}$ as $-\left(s_{\alpha_1B_1}+\dots+s_{\alpha_1B_{2M}}\right)$ and write
the vertices in \eqref{BCJ-prop1} by functions of momentums of on-shell legs.

\begin{table}\centering{\tiny\begin{tabular}{|c|c|c|c|c|c|}
                  \hline
                   & $s_{\alpha_1B_{2i+1}}\times k_{B_{2j+1}}^2$ & $s_{\alpha_1B_{2i+1}}\times k_{B_{2j}}^2$ & $s_{\alpha_1B_{2i}}\times k_{B_{2j+1}}^2$ & $s_{\alpha_1B_{2i}}\times k_{B_{2j}}^2$ \\ \hline
                  Type-A &  $\Bigl\{\begin{array}{cc}
                                       -2(i-j) & (i>j) \\
                                       0 & (i\leq j)
                                     \end{array}$
                   & $\Bigl\{\begin{array}{cc}
                       -2j & (i\geq j) \\
                       -(2i+1) &(i< j) \\
                     \end{array}$
                     & $\Bigl\{\begin{array}{cc}
                         -(2i-2j-1) & (i>j) \\
                         0 & (i\leq j)
                       \end{array}$
                       &   $\Bigl\{\begin{array}{cc}
                             -2j & (i>j) \\
                             -2i & (i\leq j)
                           \end{array}$
                         \\ \hline
                  Type-B & $\Bigl\{\begin{array}{cc}
                                       2(i-j) & (i>j) \\
                                       0 & (i\leq j)
                                     \end{array}$  & $\Bigl\{\begin{array}{cc}
                                                       2j-1 & (i\geq j) \\
                                                       2i & (i<j)
                                                     \end{array}$
                                       & $\Bigl\{\begin{array}{cc}
                                           2i-2j-1 & (i>j) \\
                                           0 & (i\leq j)
                                         \end{array}$
                                         & $\Bigl\{\begin{array}{cc}
                                             2j-1 & (i>j) \\
                                             2i-1 & (i\leq j)
                                           \end{array}$
                                            \\ \hline
                  Type-C &  0 & 1  &  0 &  1   \\
                  \hline
                \end{tabular}
}\caption{Coefficients of $J(\{B_1\})\dots J(\{B_{2M}\})$ in fundamental BCJ relation: Coefficients of the form $s_{\alpha_1B_i}\times p_{B_j}^2$ with arbitrary $i$ and $j$.}\label{t2}\end{table}

\begin{table}\centering{\tiny\begin{tabular}{|c|c|c|c|c|c|}
  \hline
    & $s_{\alpha_1B_{2i+1}}\times s_{\alpha_1B_{2j+1}}$ & $s_{\alpha_1B_{2i+1}}\times s_{\alpha_1B_{2j}}$& $s_{\alpha_1B_{2i+1}}\times s_{B_{2j+1}B_{2l+1}}$ & $s_{\alpha_1B_{2i+1}}\times s_{B_{2j}B_{2l}}$ & $s_{\alpha_1B_{2i+1}}\times s_{B_{2j+1}B_{2l}}$ \\ \hline
  Type-A & $\Bigl\{\begin{array}{cc}
             -(i-j) & (i>j) \\
             0 & (i\leq j)
           \end{array}$
    &  $\Bigl\{\begin{array}{cc}
             -j & (i\geq j) \\
             -(i+1) & (i< j)
           \end{array}$ & $\Bigl\{\begin{array}{cc}
                            -2(i-l) & (j<l<i) \\
                            0 & Otherwise
                          \end{array}$
             & $\Bigl\{\begin{array}{cc}
                 -(2i+1) & (i\leq j<l) \\
                 -2j & (j<l, j<i)
               \end{array}$
               & $\Biggl\{\begin{array}{cc}
                   -2(i-j) & (j<i<l) \\
                   -(2l-2j-1) & (j<l\leq i) \\
                   0 & Otherwise
                 \end{array}$
                 \\ \hline
  Type-B &  $\Bigl\{\begin{array}{cc}
              i-j & (i>j) \\
              0 & (i\leq j)
            \end{array}$
   & $\Bigl\{\begin{array}{cc}
       j & (i\geq j) \\
       i & (i<j)
     \end{array}$
     & $\Bigl\{\begin{array}{cc}
          2(i-l) & (j<l<i) \\
          0 & Otherwise
        \end{array}
     $  &  $\Bigl\{\begin{array}{cc}
             2i & (i\leq j<l) \\
             2j-1 & (j<l,j<i)
           \end{array}$
      &  $\Biggl\{\begin{array}{cc}
          2(i-j) & (j<i<l) \\
           (2l-2j-1) & (j<l\leq i) \\
           0 & Otherwise
         \end{array}$
       \\ \hline
  Type-C &  0 &  0 & 0  & 1  &  0 \\
  \hline
\end{tabular}}\caption{Coefficients of $J(\{B_1\})\dots J(\{B_{2M}\})$ in fundamental BCJ relation: Coefficients of the form $s_{\alpha_1B_{2i+1}}\times \dots$.}\label{t3}\end{table}

\begin{table}\centering{\tiny\begin{tabular}{|c|c|c|c|c|c|}
  \hline
   & $s_{\alpha_1B_{2i}}\times s_{\alpha_1B_{2j+1}}$ & $s_{\alpha_1B_{2i}}\times s_{\alpha_1B_{2j}}$& $s_{\alpha_1B_{2i}}\times s_{B_{2j+1}B_{2l+1}}$ & $s_{\alpha_1B_{2i}}\times s_{B_{2j}B_{2l}}$ & $s_{\alpha_1B_{2i}}\times s_{B_{2j+1}B_{2l}}$ \\ \hline
  Type-A & $\Bigl\{\begin{array}{cc}
             -(i-j-1) & (i>j+1) \\
             0 & Otherwise
           \end{array}$
    & $\Bigl\{\begin{array}{cc}
        -j & (i>j) \\
        -i & (i\leq j)
      \end{array}$
      &  $\Bigl\{\begin{array}{cc}
           -(2i-2l-1) & (j<l<i) \\
           0 & Otherwise
         \end{array}$
       & $\Bigl\{\begin{array}{cc}
           -2i & (i\leq j<l) \\
           -2j & (j<l,j<i)
         \end{array}$
         & $\Biggl\{\begin{array}{cc}
             -(2i-2j-1) & (j<i\leq l) \\
             -(2l-2j-1) & (j<l<i) \\
             0 & Otherwise
           \end{array}$
           \\ \hline
  Type-B & $\Bigl\{\begin{array}{cc}
             i-j & (i>j) \\
             0 & (i\leq j)
           \end{array}$
    & $\Bigl\{\begin{array}{cc}
        j & (i>j) \\
        i & (i\leq j)
      \end{array}$
      & $\Bigl\{\begin{array}{cc}
          2i-2l-1 & (j<l<i) \\
          0 & Otherwise
        \end{array}$
        & $ \Bigl\{\begin{array}{cc}
             2i-1 & (i\leq j<l) \\
             2j-1 & (j<l,j<i)
           \end{array}$
         & $\Biggl\{\begin{array}{cc}
             2i-2j-1 & (j<i\leq l) \\
             2l-2j-1 & (j<l<i) \\
             0 & Otherwise
           \end{array}$
           \\ \hline
  Type-C &  0 &  0 &  0 &  1 & 0  \\
  \hline
\end{tabular}}\caption{Coefficients of $J(\{B_1\})\dots J(\{B_{2M}\})$ in fundamental BCJ relation: Coefficients of the form $s_{\alpha_1B_{2i}}\times \dots$.}\label{t4}\end{table}

After these steps, we can read off the coefficient of $J(\{B_1\})\dots J(\{B_{2M}\})$ explicitly. They are shown in tables \ref{t2}, \ref{t3}, \ref{t4}.
The columns of tables \ref{t2}, \ref{t3}, \ref{t4}, except for the second column of table \ref{t3} and the first column of table \ref{t4}, are canceled out.
The sum of the second column of table \ref{t3} is given as
\bea
\Biggl\{
  \begin{array}{cc}
    0 & (i\geq j) \\
    -1 & (i<j) \\
  \end{array},
\eea
while, the sum of the first column of table \ref{t4} is given as
\bea
\Biggl\{\begin{array}{cc}
    1 & (i> j) \\
    0 & (i\leq j) \\
  \end{array}.
\eea
Since $s_{\alpha_1\beta_{2i+1}}\times s_{\alpha_1\beta_{2j}}$ and $s_{\alpha_1\beta_{2i}}\times s_{\alpha_1\beta_{2j+1}}$ can be related by $i\Leftrightarrow j$,
we should interchange $i$ and $j$ in the first column of table \ref{t4}. Then we can see these two nonzero contributions cancel with each other.
Therefore, all the contributions  of divisions with $M>1$ at last must vanish.

For division with $M=1$, the ordered set $\{\beta\}$ \cgg{are}{is} only divided into two ordered subsets. In this case, we only need to consider the terms of $(p_1^2)^0$ in diagrams shown in Fig. \ref{bcj_sub2} (which is the first term of the second line of \eqref{BCJ-prop1}) with all the possible nontrivial divisions $\{\beta\}\rightarrow\{B_1\}\{B_2\}$. The sum of these terms precisely gives the right hand side of the fundamental BCJ relation \eqref{off-shell-BCJ}.

\section{General KK, BCJ relations, minimal-basis expansion and formulations of total amplitudes}

Having proven the $U(1)$-decoupling identity and fundamental BCJ relation in non-linear sigma model, let us now extend these relations to more general cases.
In this section, we first state that the general KK and BCJ relations as well as minimal-basis expansion are all satisfied by color-ordered tree amplitudes.
Then we will show that tree-level total amplitudes satisfy  DDM form of color decomposition and KLT relation\footnote{We emphasize that the consequent relations that will be derived in this section are all for on-shell amplitudes. General KK and BCJ relations for off-shell currents will be discussed in future work.}. All these discussions are parallel within Yang-Mills theory, thus we will only present the main points of the statements. Details can be found in the works \cite{Ma:2011um}, \cite{Chen:2011jxa}, \cite{DelDuca:1999rs} and \cite{Du:2011js}.

\subsection{General KK, BCJ relations and minimal-basis expansion}

{\emph {General KK and BCJ relations}}

KK relation and general BCJ relation can be considered as extensions of $U(1)$-decoupling identity and fundamental BCJ relation. In non-linear sigma model, KK relation for $2m$-point amplitudes is given as
\bea
\Sl_{\sigma\in OP(\{\alpha_1,\dots,\alpha_{r}\}\bigcup\{\beta_1,\dots,\beta_{s}\})}A(1,\{\sigma\},2m)=(-1)^rA(1,\{\beta\},2m,\{\alpha\}^T),~~\label{on-shell-gen-KK}
\eea
where $r+s=2m-2$. General BCJ relation is given as
\bea
\Sl_{\sigma\in OP(\{\alpha_1\dots \alpha_r\}\bigcup\{\beta_1,\dots,\beta_{s}\})}\Sl_{l=1}^r\Sl_{\xi_{\sigma_i}<\xi_{\alpha_l}}s_{\alpha_l\sigma_i}A(1,\{\sigma\},2m)=0.\label{on-shell-gen-BCJ}
\eea
From \eqref{on-shell-gen-KK} and \eqref{on-shell-gen-BCJ}, we can see, if there is only one element in $\{\alpha\}$, the relations turns back to the $U(1)$-decoupling identity \eqref{on-shell-U(1)} and the fundamental BCJ relation \eqref{on-shell-BCJ} with $2m\rightarrow \beta_{2m}$.

In principle, one can follow the similar steps in sections 3 and 4 to prove the general KK , BCJ relations \eqref{on-shell-gen-KK}, \eqref{on-shell-gen-BCJ} for off-shell currents and then take on-shell limits to get the relations among color-ordered on-shell amplitudes. However, it is not easy to generalize the off-shell KK and BCJ relations in this way. This is because there are nontrivial products of sub-currents on the right hand side of the relations. When there are more elements in $\{\alpha\}$ set, the forms of the right hand side may containing both divisions of $\{\alpha\}$ set and divisions of $\{\beta\}$ set. Thus the formulations may become highly complicated.

Fortunately, once we know the fundamental BCJ relation \eqref{on-shell-BCJ} in addition with cyclic symmetry \eqref{Cyclic symmetry}, we have another way to prove the on-shell general KK and BCJ relations. This method was firstly proposed in \cite{Ma:2011um} where general KK and BCJ relations in Yang-Mills theory are generated by so-called \emph{primary relations}. The main point is that once the amplitudes satisfy \emph{a)cyclic symmetry} as well as \emph{b)fundamental BCJ relation}, all the general KK and BCJ relations can be reexpressed as linear combinations of a set of fundamental BCJ relations, and thus the general KK, BCJ relations must hold.
Though the discussions in \cite{Ma:2011um} was firstly found by monodromy relations in string theory, as stated in \cite{Ma:2011um}, all these arguments can be extended to field theory. Since the fundamental BCJ relation\eqref{on-shell-BCJ} in non-linear sigma model has the same form within Yang-Mills theory, all the steps in \cite{Ma:2011um} are also valid in non-linear sigma model. Thus the KK and BCJ relations must be satisfied by color-ordered tree amplitudes in non-linear sigma model.
Details of this proof can be found in \cite{Ma:2011um}.

{\emph {Minimal-basis expansion}}

Since KK and general BCJ relations are both satisfied by even-point color ordered tree amplitudes. We are ready now for reduce the number of independent even-point color ordered tree amplitudes as in Yang-Mills theory. Apparently, one can use KK relation in addition with cyclic symmetry to reduce the number of independent $2m$-point amplitudes to $(2m-2)!$. As in Yang-Mills theory, BCJ relations provide further constraints. One can use general BCJ relations to express the amplitudes in KK basis by only $(2m-3)!$ independent amplitudes. The explicit formation of minimal-basis expansion is Eq. (4.22) in \cite{Bern:2008qj} with $2m$ external legs. One can follow the same recursive procedure that given by section 4 of the paper \cite{Chen:2011jxa} to prove the minimal-basis expansion, because we have the \emph{general BCJ relation} \eqref{on-shell-gen-BCJ} of the same form within Yang-Mills theory.

\subsection{Formulations of total amplitudes}

In Yang-Mills theory, amplitude relations imply various formations of total amplitudes. As we have discussed, in non-linear sigma model, event-point color ordered tree amplitudes satisfy KK and BCJ relations, which have the same formations within Yang-Mills theory. Thus we expect that the total amplitudes can have the same expressions within Yang-Mills theory. Particularly, the total amplitudes should satisfy DDM color decomposition as well as KLT relation\footnote{KLT relation in Yang-Mills theory was suggested in \cite{Bern:1999bx} and the general proof can be found in\cite{Du:2011js}.}.

{\emph {DDM form of  color decomposition}}

An immediate result of KK relation is that the total amplitudes satisfy Del Duca-Dixon-Maltoni(DDM) form of color decomposition which was firstly proven in Yang-Mills theory\cite{DelDuca:1999rs}
\bea
M(1,\dots,2m)=\Sl_{\sigma\in S_{2m-2}}f^{a_1a_{\sigma_2}a_{i_1}}\dots f^{a_{i_{2m-3}}a_{\sigma_{2m-1}}a_{2m}} A(1,\sigma,2m).\label{DDM form}
\eea
The main points to prove DDM form of color decomposition are \emph{a) KK relations}\eqref{on-shell-gen-KK} and \emph{b) the following relations between trace factors and color factors in DDM form}
\bea
f^{a_1a_{\sigma_2}a_{i_1}}\dots f^{a_{i_{2m-3}}a_{\sigma_{2m-1}}a_{2m}}=\Tr(T^{1}[T^{\sigma_2},[...,[T^{\sigma_{2m-1}},T^{2m}]...]]).
\label{c-tau-relation}\eea
We can express any color-ordered amplitude in \eqref{Trace form} by KK relation, and collect the color coefficients of each amplitude in KK basis. Using the above relation between traces and the color factors in DDM form, we can prove the DDM form of color decomposition\eqref{DDM form}.
Details of the proof can be found in \cite{DelDuca:1999rs}.

{\emph {KLT relation}}

Another result is Kawai-Lewellen-Tye(KLT) relation\cite{Kawai:1985xq}. In non-linear sigma model, total amplitudes can be expressed in terms of products of two color-ordered tree amplitudes $A$ and $\W A$, where $A$ denote the color-ordered tree amplitudes in non-linear sigma model and $\W A$ denote the color-ordered tree amplitudes of scalar with cubic vertex $f^{abc}$. As in Yang-Mills theory, the KLT relation has many formations\cite{BjerrumBohr:2010ta,BjerrumBohr:2010yc}.

For example the formulation {\emph{manifests $(2m-2)!$ symmetries}} is given as
\bea
M(1,2,\dots,2m)=\Sl_{\gamma,\phi\in S_{2m-2}}{A(2m,\gamma,1)S[\gamma|\phi]\W A(1,\phi,2m)\over s_{12\dots(2m-1)}}.~~~~~\label{(2m-2)!KLT}
\eea
This relation can be proved by following the same steps within the subsection 6.3 of the paper \cite{Du:2011js}. This is because that the two critical points-\emph{the DDM color decomposition} and \emph{the generalized BCJ relation for color scalar theory}-are all satisfied.

Another formulation which {\emph{manifests $(2m-3)!$ symmetries}} is given as
\bea
M(1,2,\dots,2m)=(-1)\Sl_{\gamma,\phi\in S_{2m-3}}A(1,\gamma,2m-1,2m)S[\phi|\gamma]_{1}\W A(2m-1,2m,\phi,1),~~~~\label{(2m-3)!KLT}
\eea
or equivalently
\bea
M(1,2,\dots,2m)=(-1)\Sl_{\gamma,\phi\in S_{2m-3}}A(1,\gamma,2m-1,2m)S[\gamma|\phi]_{p_{n-1}}\W A(1,2m-1,\phi,2m).
\eea
This formulation seems not easy to prove along the same line in Yang-Mills theory (See section 6.1 of \cite{Du:2011js}), because the
boundary behavior of the amplitudes of non-linear sigma model is not good enough. However, we also expect that the $(2m-3)!$ formulation have
the same form within Yang-Mills theory. In this paper, we just take  the four-point KLT relation as an example
\bea
M(1,2,3,4)=-A(1,2,3,4)s_{21}\W A(4,2,1,3).
\eea
To prove this relation, we express $\W A(4,2,1,3)$ explicitly by Feynman rules in color scalar theory. Thus the right hand side is expressed as
\bea
-A(1,2,3,4)s_{21}\left[{f^{13e}f^{e42}\over s_{13}}+{f^{21e}f^{e34}\over s_{12}}\right].
\eea
Using antisymmetry of $f^{abc}$ as well as four-point BCJ relation  $s_{12}A(1,2,3,4)+(s_{12}+s_{23})A(1324)=0$ which have been proven in the previous sections,
we reexpress the right hand side as
\bea
f^{12e}f^{e34}A(1,2,3,4)+f^{13e}f^{e24}A(1,3,2,4).
\eea
This is just the DDM form of color decomposition of four-point total tree amplitude. Thus the four-point KLT relation manifest $(4-3)!=1$ symmetry is proved.
We leave the general proof of this formula for future discussion.

Though KLT relation was suggested in gravity and then in Yang-Mills theory,
it is not surprising that the double-copy formula can also exist in a scalar theory such as non-linear sigma model. An example for KLT relation of scalar amplitudes can be found in bosonic string theory where the closed string tachyon amplitudes at tree level can be expressed by double copy of open string tachyon amplitudes\cite{Kawai:1985xq}. Actually, the non-linear sigma model also have the similar double-copy structure when considering the color part and the kinematic part as the two copies.

\section{Conclusion}

In this work, we have discussed the tree-level amplitude relations in non-linear sigma model. We have proven the off-shell version of $U(1)$ identity and fundamental BCJ relation under Cayley parametrization. After taking on-shell limits, we got the $U(1)$-decoupling identity and the fundamental BCJ relation for on-shell amplitudes. We stated that the general KK and BCJ relations were also satisfied by even-point tree amplitudes in non-linear sigma model.
Two consequent results of KK and BCJ relations were given as the minimal-basis expansion for color-ordered amplitudes and KLT relation for total amplitudes. Though the procedure of proof in this work seems complicated, the relations are quite consistent with the color algebra. We hope these results can be useful in particle phenomenology. The algebraic interpretation of these relations and the dual decompositions of amplitudes deserve further work.

\appendix

\section{Convention of notation}

\begin{figure}[h!]
  \centering
 \includegraphics[width=1\textwidth]{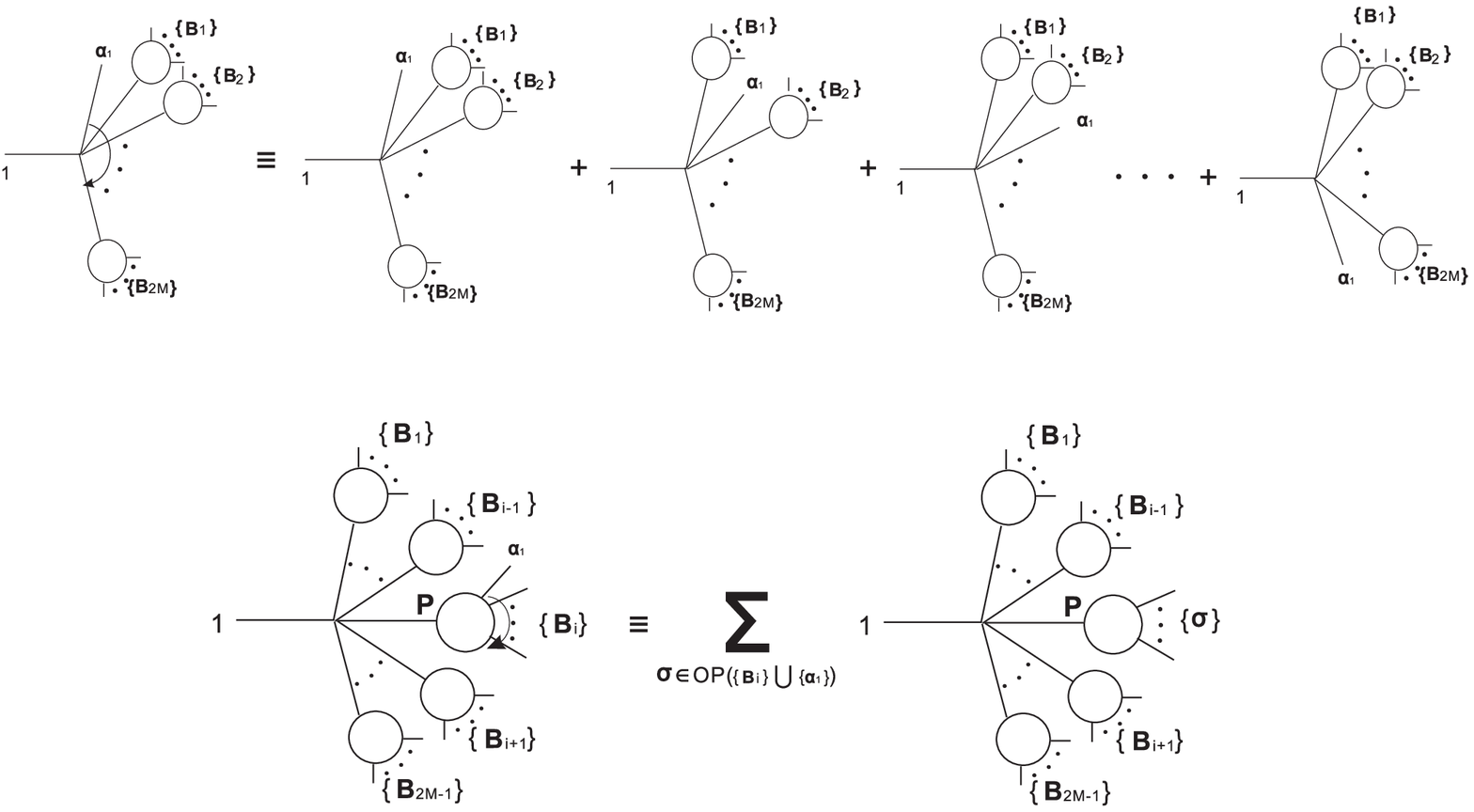}
 \caption{Convention in section 3}\label{convention3}
\end{figure}

\begin{figure}[h!]
  \centering
 \includegraphics[width=1\textwidth]{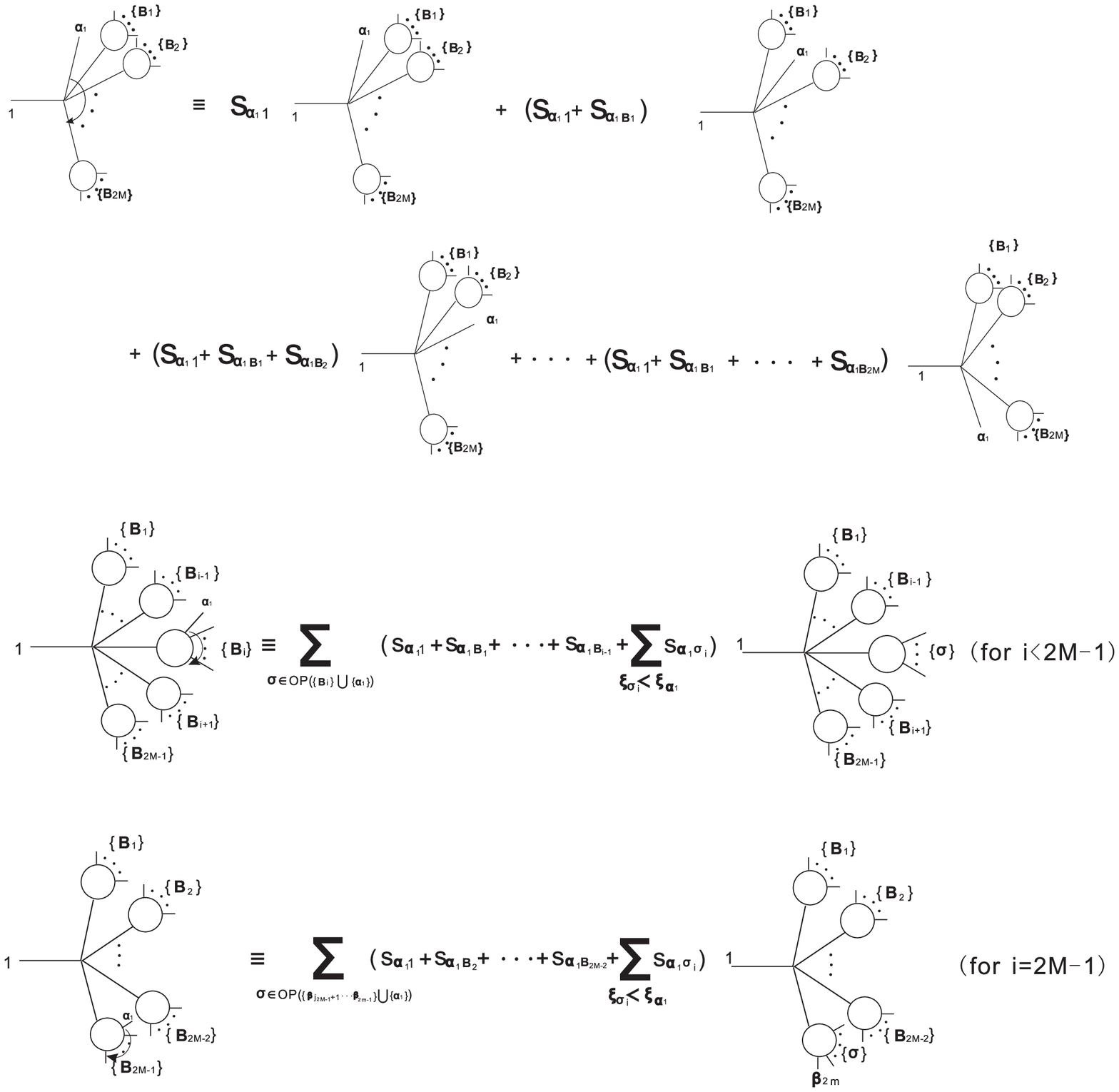}
 \caption{Convention in section 4}\label{convention4}
\end{figure}

In this paper, we use a diagram containing a curved arrow line to denote sum of diagrams for short. Since we encounter similar structures when considering
$U(1)$ identity and fundamental BCJ relation, we only use the same diagrams expressions but let the curved arrow line have different meanings for convenience.
The meaning of curved arrow line for section 3 and section 4 are given by Fig. \ref{convention3} and Fig. \ref{convention4} respectively.

\section{Eight-point diagrams}
\begin{figure}[h!]
  \centering
 \includegraphics[width=0.83\textwidth]{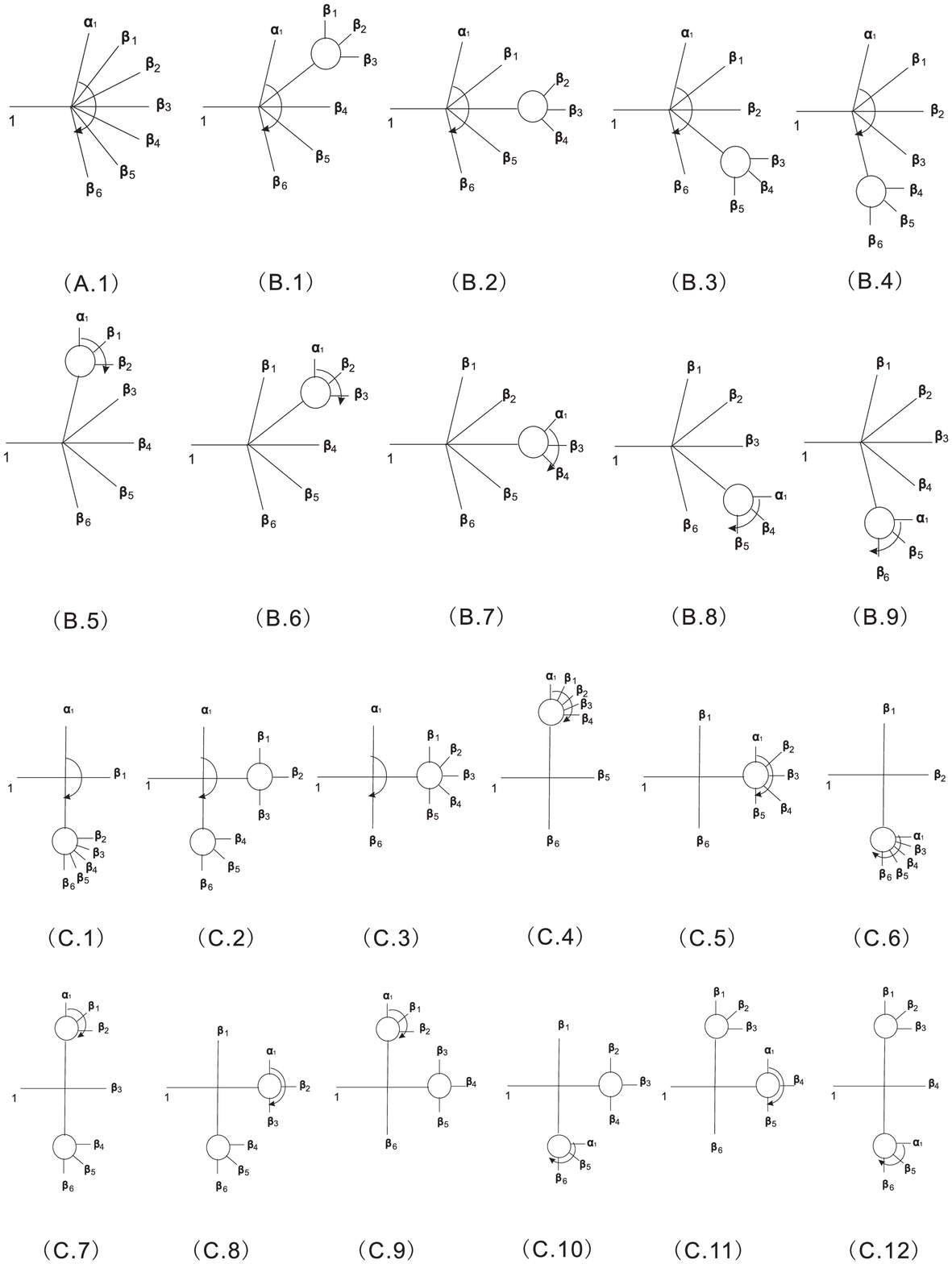}
 \caption{Diagrams for eight-point U(1) identity(with curved arrow line defined by Fig. 3) or fundamental BCJ relation(with curved arrow line defined by Fig. 4) } \label{8pt_diagrams}
\end{figure}

The left hand side of eight-point $U(1)$ identity and eight-point fundamental BCJ relation can be expressed by Fig.  \ref{8pt_diagrams} with the convention of notation defined by Fig. \ref{convention3} and Fig. \ref{convention4}.
%

\section{Diagrams contribute to $J(\{B_1\})J(\{B_2\})\dots J(\{B_{2M}\})$}
\begin{figure}[h!]
  \centering
 \includegraphics[width=0.83\textwidth]{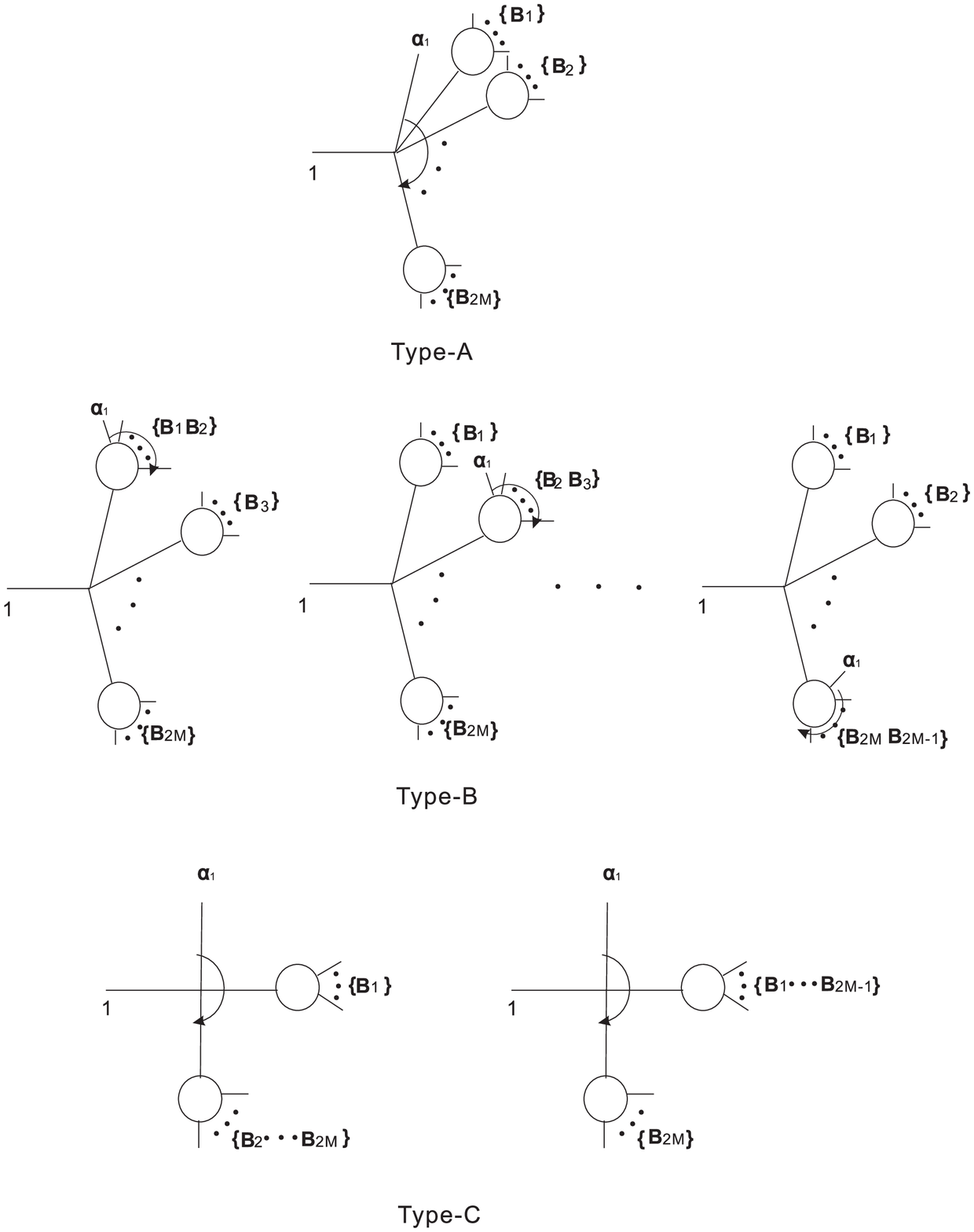}
 \caption{The three types of diagrams contribute to $J_{B_1}J_{B_2}\dots J_{B_{2M}}$. } \label{gen}
\end{figure}

The diagrams contribute to $J(\{B_1\})J(\{B_2\})\dots J(\{B_{2M}\})$ in $U(1)$ identity and fundamental BCJ relation are given by Fig. \ref{gen}.
%

\subsection*{Acknowledgements}
Y. J. Du is pleased to thank Prof. Yong-Shi Wu for helpful discussions. He would also like to thank the University of Utah for hospitality.
Y. J. Du is supported in part by the NSF of China Grant No.11105118, China Postdoctoral Science Foundation No.2013M530175
and the Fundamental Research Funds for the Central Universities of Fudan University No.20520133169. The research of Gang Chen has been
supported in parts by the Jiangsu Ministry of Science and Technology
under contract~BK20131264 and by the Swedish Research Links programme of the Swedish Research Council (Vetenskapsradets generella villkor) under contract~348-2008-6049. Gang Chen  also acknowledges
985 Grants from the Ministry of Education, and the
Priority Academic Program Development for Jiangsu Higher Education Institutions (PAPD).


\end{document}